\documentclass[12pt]{article}
\pdfoutput=1
\usepackage[a4paper]{geometry}
\usepackage{jheppub, amsmath,amssymb,amsfonts,amsxtra, mathrsfs, makeidx,graphics,graphicx,amsthm,epsfig, ytableau,bm,longtable,float, color,tikz,mathtools,xfrac,footnote,rotating, lscape}
\usetikzlibrary{positioning,shapes}
\usetikzlibrary{chains}
\usetikzlibrary{arrows,fit,decorations.pathreplacing}
\tikzstyle{every picture}+=[remember picture]
\tikzstyle{na} = [baseline=-.5ex]

\usepackage{bbm}

\newcommand{\eg}{\textit{e.g.}}

\newcommand{\ie}{\textit{i.e.}}

\numberwithin{equation}{section}

\newcommand{\nn}{\nonumber}

\newcommand{\be}{\begin{equation}} \newcommand{\ee}{\end{equation}}
\newcommand{\bea}{\begin{equation} \begin{aligned}} \newcommand{\eea}{\end{aligned} \end{equation}}

\def\hat{\widehat}

\def\rt2{\sqrt{2}}

\def\det{\mathop{\rm det}}

\def\Tr{\mathop{\rm Tr}}

\def\CH{{\cal H}}

\def\CN{{\cal N}}

\def\CS{{\cal S}}

\def\1{{\ds 1}}

\newcommand{\cG}{\mathcal{G}}

\newcommand{\cN}{\mathcal{N}}

\newcommand{\cR}{\mathcal{R}}

\newcommand{\cW}{\mathcal{W}}

\newcommand{\bZ}{\mathbb{Z}}

\def\repa{\raise4pt\hbox{$\square$}\mkern-14mu\raise-4pt\hbox{$\square$}}
\def\repab{\overline{\raise4pt\hbox{$\square$}\mkern-14mu\raise-4pt\hbox{$\square$}\mkern-1mu}}

\def\smileface{\ensuremath{\hbox{\large$\bigcirc$}\mkern-15mu\raise-1pt\hbox{\scriptsize$\smallsmile$}%
\mkern-10mu\raise4pt\hbox{..}\mkern4mu}}
\def\frownface{\ensuremath{\hbox{\large$\bigcirc$}\mkern-15mu\raise-1pt\hbox{\scriptsize$\smallfrown$}%
\mkern-10mu\raise4pt\hbox{..}\mkern4mu}}

\def\node#1#2{\overset{#1}{\underset{#2}{\circ}}}

\newcommand{\ba}{\begin{array}}
\newcommand{\ea}{\end{array}}
\newcommand{\bi}{\begin{itemize}}
\newcommand{\ei}{\end{itemize}}
\def\vec#1{\bm{#1}}
\def\bea#1\eea{\allowdisplaybreaks \begin{align}#1\end{align}}
 \newcommand{\ben}{\begin{enumerate}}
\newcommand{\een}{\end{enumerate}}
\newcommand{\bean}{\begin{eqnarray*}}
\newcommand{\eean}{\end{eqnarray*}}
\newcommand{\eref}[1]{(\ref{#1})}

\newcommand{\PE}{\mathop{\rm PE}}
\newcommand{\PL}{\mathop{\rm PL}}

\newcommand{\BC}{\mathbb{C}}

\newcommand{\BZ}{\mathbb{Z}}

\newcommand{\BH}{\mathbb{H}}

\newcommand{\comment}[1]{}
\newcommand{\diag}{\mathrm{diag}}

\title{The Moduli Space of Instantons on an ALE Space from $3d$ $\CN=4$ Field Theories}
\author{Noppadol Mekareeya}

\affiliation{Theory Division, Physics Department, CERN, \\CH-1211, Geneva 23, Switzerland}
\emailAdd{noppadol.mekareeya@cern.ch}
\abstract{The moduli space of instantons on an ALE space is studied using the moduli space of $\CN=4$ field theories in three dimensions. For instantons in a simple gauge group $G$ on $\BC^2/\BZ_n$, the Hilbert series of such an instanton moduli space is computed from the Coulomb branch of the quiver given by the affine Dynkin diagram of $G$ with flavour nodes of unitary groups attached to various nodes of the Dynkin diagram.  We provide a simple prescription to determine the ranks and the positions of these flavour nodes from the order of the orbifold $n$ and from the residual subgroup of $G$ that is left unbroken by the monodromy of the gauge field at infinity.   For $G$ a simply laced group of type $A$, $D$ or $E$, the Higgs branch of such a quiver describes the moduli space of instantons in projective unitary group $PU(n) \cong U(n)/U(1)$ on orbifold $\BC^2/\widehat{G}$, where $\widehat{G}$ is the discrete group that is in McKay correspondence to $G$.  Moreover, we present the quiver whose Coulomb branch describes the moduli space of $SO(2N)$ instantons on a smooth ALE space of type $A_{2n-1}$ and whose Higgs branch describes the moduli space of $PU(2n)$ instantons on a smooth ALE space of type $D_{N}$.}

\preprint{
{\small
\begin{flushright}
CERN-PH-TH-2015-204
\end{flushright}
}
}

\begin{document}
\maketitle

\section{Introduction}
The realisation of the moduli space of instantons using string theory and supersymmetric field theory has long been studied in the literature.  It was pointed out in \cite{Witten:1995gx, Douglas:1995bn} that the ADHM construction \cite{Atiyah:1978ri}, an algebraic prescription to construct instanton solutions for classical gauge groups on flat space, can be realised from a system of D$p$-branes inside D$(p+4)$-branes (possibly with the presence of an orientifold plane).  In this system, the former dissolve into instantons for the worldvolume gauge fields of the latter.  For the worldvolume gauge theory on the D$p$-branes, which has 8 supercharges, the Higgs branch of the moduli space can then be identified with the moduli space of instantons in the theory on the worldvolume on the D$(p+4)$ branes.  Indeed, the hyperK\"ahler quotient of such a Higgs branch can be identified with that of the moduli space of instantons \cite{Hitchin:1986ea}.   

A similar method is also available for instantons on an asymptotically locally Euclidean (ALE) space.  The analogue of the ADHM construction for instantons in a unitary gauge group on an ALE space was proposed by Kronheimer and Nakajima (KN) \cite{kronheimer1990yang}.  Such a result was later generalised to instantons in a general classical gauge group on an ALE space of type $A$ or $D$ by several authors, \eg~ \cite{Douglas:1996sw, deBoer:1996mp, Porrati:1996xi, Intriligator:1997kq, Hanany:1997gh, Hanany:1999sj}, using string theoretic constructions.  

Both ADHM and KN constructions exist only for instantons in classical gauge groups.  Over the past few years, there has been a lot of progress in the study of instantons in exceptional gauge groups on flat space.  For example, the instanton partition functions for exceptional gauge groups $E_{6,7,8}$ can be derived \cite{Gaiotto:2012uq} by computing of superconformal indices \cite{Romelsberger:2005eg,Kinney:2005ej,Gadde:2011uv} of theories of class $\CS$, arising from compactifying M5-branes on a Riemann sphere with appropriate punctures \cite{Gaiotto:2009we}; by using the blow-up equations of \cite{Nakajima:2003pg,Nakajima:2005fg} as in \cite{Keller:2012da}; or by computing and extrapolating the generating function of holomorphic functions on the Higgs branch, known as the Higgs branch Hilbert series, of certain supersymmetric field theories as in \cite{Benvenuti:2010pq, Keller:2011ek, Hanany:2012dm}.  

Another method that has been proven to be fruitful for the study of instanton moduli spaces is to make use of the Coulomb branch of certain three dimensional field theories with 8 supercharges. For one instanton in a simply laced gauge group (\ie~ of type $A$, $D$ or $E$) on flat space, such $3d$ gauge theories were proposed in \cite{Intriligator:1996ex}.  Such results were later extended to higher instanton numbers and to instantons on ALE spaces by \cite{deBoer:1996mp, Porrati:1996xi, Hanany:1999sj}.  Due to the method discovered in \cite{Cremonesi:2013lqa}, it became possible to compute the generating function of holomorphic functions on the Coulomb branch, known as the Coulomb branch Hilbert series, of a large class of $3d$ $\CN=4$ gauge theories (see also \cite{Nakajima:2015txa} for a recent mathematical development).  It was indeed shown that this method can be applied to study the moduli space of instantons in the ADE gauge groups on flat space \cite{Cremonesi:2013lqa, Cremonesi:2014vla}.  The generalisation to the non-simply laced gauge groups (\ie~ of types $B$, $C$, $F$ and $G$) was developed recently in \cite{Cremonesi:2014xha}.  In this reference, it was proposed that the moduli space of instantons in any simple gauge group $G$, including non-simply laced ones, on flat space can be realised from the Coulomb branch of the quiver given by the over-extended Dynkin diagram of $G$.  For $G$ a simply laced group, such a quiver represents a $3d$ $\CN=4$ gauge theory with a known Lagrangian description, and this is indeed the main reason why the method of \cite{Cremonesi:2013lqa} can be successfully applied to study such a moduli space.  On the other hand, the Lagrangian description for non-simply laced Dynkin diagrams is not currently known, due to the multiple laces that are present in these diagrams. Nevertheless, it is still possible to compute the Coulomb Hilbert series such quivers using a simple prescription presented in \cite{Cremonesi:2014xha}.

The Hilbert series is, mathematically speaking, a character of the global symmetry group of the ring of holomorphic functions on the moduli space of the supersymmetric gauge theory. It provides useful information about the moduli space, namely the group theoretic properties of the generators of the moduli space and of the relations between them. Important properties of the theory, such as the global symmetry enhancement, can be studied using this approach.  For moduli spaces of $k$ pure Yang-Mills instantons, the Hilbert series is also the five-dimensional ($K$-theoretic) $k$ instanton partition function of \cite{Nekrasov:2002qd, Nakajima:2003pg, Nekrasov:2004vw, Nakajima:2005fg, Keller:2011ek, Nakajima:2015txa}.  Moreover, it was recently realised that the Coulomb branch Hilbert series of $3d$ $\CN = 4$ ``good" or ``ugly" theories \cite{Gaiotto:2008ak} with a Lagrangian description \cite{Cremonesi:2013lqa} can be obtained as a limit of the superconformal index of the theory\footnote{The derivation of this limit in \cite{Razamat:2014pta} is for a $U(N)$ gauge group, but it can be easily generalised to any gauge group and matter content.} \cite{Razamat:2014pta}.    Indeed, some of the theories of our interest in this paper, in particular those involving the simply-laced Dynkin diagrams, have a Lagrangian description and the standard formula \cite{Kim:2009wb, Imamura:2011su, Krattenthaler:2011da, Kapustin:2011jm} for the superconformal index can be written down.  On the other hand, the theories involving non-simply laced Dynkin diagrams have no known Lagrangian description and the standard formula for the superconformal index is not available. The prescription for dealing with multiple laces proposed in \cite{Cremonesi:2014xha} is a natural generalisation of the result obtained in \cite{Cremonesi:2013lqa} for Lagrangian theories.  Such an approach allows us to study in a uniform way the moduli spaces of instantons of all simple Lie groups. It is currently unclear how to extend this to the full superconformal index.

The main goal of is paper is to study $3d$ $\CN=4$ field theories whose Higgs and/or Coulomb branches can be identified with the moduli space of instantons on an ALE space. We divide our study into two parts, namely for an orbifold singularity and for a smooth ALE space.  Let us first discuss our findings on the case of orbifold singularities.  We propose, along the line of  \cite{Cremonesi:2014xha}, a simple field theory description of the moduli space of $G$ instantons on orbifold $\BC^2/\BZ_n$, for any simple group $G$.  Such a description involves the Coulomb branch of the quiver given by the affine Dynkin diagram of $G$ with flavour nodes of unitary groups attached to various nodes of the Dynkin diagram.  We provide a simple prescription to determine the ranks and the positions of these flavour nodes from the order of the orbifold $n$ and from the residual subgroup of $G$ that is left unbroken by the monodromy at infinity.  For $G$ a simply laced group of type $A$, $D$ or $E$, the Higgs branch of such a quiver describes the moduli space of instantons in projective unitary group\footnote{The projective unitary group $PU(n)$ is defined as $PU(n) \cong U(n)/U(1)$.  The Lie algebra of which is isomorphic to that of special unitary group $SU(n)$.} $PU(n)$ on $\BC^2/\widehat{G}$, where $\widehat{G}$ is the discrete group that is in McKay correspondence to $G$. The exchange of the instanton gauge group and the orbifold group in the Higgs and the Coulomb branches is in accordance with \cite{deBoer:1996mp, Porrati:1996xi, Hanany:1999sj, Dey:2013nf}.   We use the Hilbert series as a tool to study such moduli spaces of instantons and find several interesting new features.  In the second part of the paper, we discuss the moduli space of instantons on a smooth ALE space.  We focus on a $3d$ $\CN=4$ gauge theory whose Coulomb branch describes the moduli space of $SO(2N)$ instantons on $\widetilde{\BC^2/\BZ_{2n}}$ with a particular monodromy at infinity.  In several special cases, we compare the results with those studied in \cite{Tachikawa:2014qaa} and find an agreement.  As proven in \cite{kronheimer1990yang}, the Higgs branch of such a gauge theory describes the moduli space of $PU(2n)$ instantons on smooth ALE space $\widetilde{\BC^2/\widehat{D}_{N}}$.

 The paper is organised as follows. In section \ref{sec:singular}, we focus on instantons on a singular orbifold.  As a warm-up exercise, we review the Higgs and the Coulomb branches of the $3d$ $\CN=4$ gauge theory that describes the moduli space of $PU(N)$ instantons on $\BC^2/\BZ_n$ in section \ref{sec:SUNonC2Zn}.  Extending the previous result, we then propose the field theories whose Coulomb branch is the moduli space of instantons in a general simple gauge group on $\BC^2/\BZ_n$ in section \ref{gensimGC2Zn}.  As a by-product, the moduli spaces of $PU(n)$ instantons on $\BC^2/\widehat{D}_{N}$ and $\BC^2/\widehat{E}_{6,7,8}$ are discussed in the same section.  Several explicit examples are given in section \ref{sec:examples} and there the Hilbert series are computed for a number of special cases to support our proposal in the preceding section. In section \ref{sec:smoothALE}, we focus on instantons on a smooth ALE space.   We propose a theory whose Coulomb branch describes the moduli space of $SO(8)$ instantons on smooth ALE space $\widetilde{\BC^2/\BZ_2}$ in section \ref{sec:SO8smoothC2Z2} and generalise this result to $SO(2N)$ instantons on $\widetilde{\BC^2/\BZ_{2n}}$ in section \ref{sec:SO2NC2Z2n}.  For simplicity, we focus only on the monodromy at infinity such that $SO(2N)$ is broken to $SO(2N-4) \times SO(4)$.  In section \ref{sec:SU2nonsmoothC2DN}, we study the Higgs branch of the aforementioned quiver and conjecture that it can be identified with the moduli space of $PU(2n)$ instantons on $\widetilde{\BC^2/\widehat{D}_{N}}$ with monodromy at infinity such that $PU(2n)$ is left unbroken.  We close the paper with a discussion of our results and present some open problems in section \ref{sec:conclude}.




\section{Instantons in gauge group $G$ on orbifold $\BC^2/\Gamma$} \label{sec:singular}
In this section, we study the moduli space of $G$ instantons on orbifold $\BC^2/\BZ_n$, for any simple group $G$.  As a warm-up exercise, we discuss the case of $G=PU(N)$ and $\Gamma=\BZ_n$ in section \ref{sec:SUNonC2Zn}.  Many results on this topic have been studied in several papers, \eg~ \cite{kronheimer1990yang, Douglas:1996sw, Bianchi:1996zj, deBoer:1996mp, Porrati:1996xi, Fucito:2004ry, Cherkis:2008ip, Witten:2009xu, Cherkis:2009jm, Dey:2013fea, Nakajima:2015txa}, in the past.  The purpose of this section is to introduce necessary concepts and set-up the notation.  In section \ref{gensimGC2Zn}, we then extend this result to the moduli space of $G$ instantons on $\BC^2/\BZ_n$ for a general simple gauge group $G$, and to the moduli spaces of $PU(n)$ instantons on $\BC^2/\widehat{D}_{N}$ and $\BC^2/\widehat{E}_{6,7,8}$.

\paragraph{Notation.}  In the quiver diagrams in this paper, we use a circular node to denote a gauge group, a rectangular node to denote a flavour symmetry, and a line connecting two nodes to denote a bi-fundamental hypermultiplet between the two groups.  Unless stated otherwise, the node labelled by number $r$ denotes the unitary group $U(r)$.

\subsection{$G=PU(N)$ and $\Gamma=\BZ_n$} \label{sec:SUNonC2Zn}
We start by discussing moduli space of $PU(N)$ instantons on $\BC^2/\BZ_n$. The instanton configuration is specified by two pieces of information, namely the monodromy of the gauge field at infinity and the monodromy at the origin of $\BC^2/\BZ_n$ \cite{Witten:2009xu}.  With these data specified, the moduli space of such instantons are isomorphic to the {\it Higgs branch} of a $3d$ $\CN=4$ gauge theory whose quiver description is given by the flavoured affine $A_{n-1}$ quiver diagram, known as the Kronheimer-Nakajima (KN) quiver \cite{kronheimer1990yang}:
\bea 
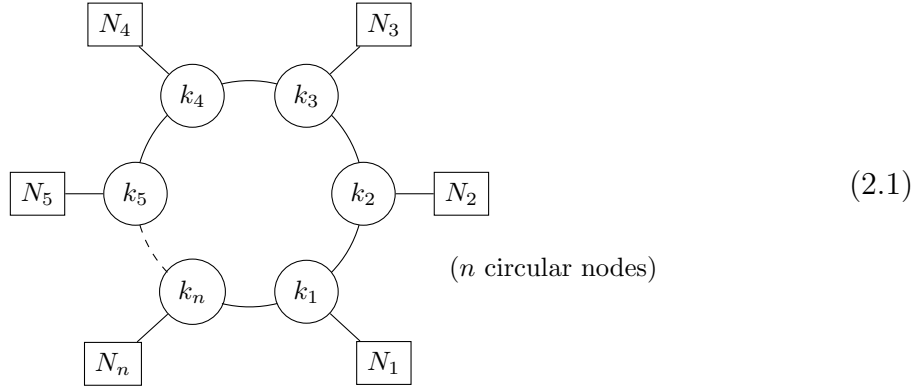
\label{KN}
\begin{tikzpicture}[baseline, align=center,node distance=0.5cm]
\def \n {6}
\def \radius {1.5cm}
\def \margin {16} 
\foreach \s in {1,...,5}
{
  \node[draw, circle] (\s) at ({360/\n * (\s - 2)}:\radius) {{\footnotesize $k_{\s}$}};
  \draw[-, >=latex] ({360/\n * (\s - 3)+\margin}:\radius) 
    arc ({360/\n * (\s - 3)+\margin}:{360/\n * (\s-2)-\margin}:\radius);
}
\node[draw, circle] (last) at ({360/3 * (3 - 1)}:\radius) {{\footnotesize $k_n$}};
\draw[dashed, >=latex] ({360/6 * (5 -2)+\margin}:\radius) 
    arc ({360/6 * (5 -2)+\margin}:{360/6 * (5-1)-\margin}:\radius);
    \node[draw, rectangle,  below right= of 1] (f1) {{\footnotesize $N_1$}};
\node[draw, rectangle, right= of 2] (f2) {{\footnotesize $N_2$}};
\node[draw, rectangle, above right= of 3] (f3) {{\footnotesize $N_3$}};
\node[draw, rectangle, above left= of 4] (f4) {{\footnotesize $N_4$}};
\node[draw, rectangle,  left= of 5] (f5) {{\footnotesize $N_5$}};
\node[draw, rectangle,  below left= of last] (f6) {{\footnotesize $N_n$}};
\draw[-, >=latex] (1) to (f1);
\draw[-, >=latex] (2) to (f2);
\draw[-, >=latex] (3) to (f3);
\draw[-, >=latex] (4) to (f4);
\draw[-, >=latex] (5) to (f5);
\draw[-, >=latex] (last) to (f6);
\node[draw=none] at (4,-1) {{\footnotesize ($n$ circular nodes)}};
\end{tikzpicture}
\eea
where
\bea
N= N_1 + N_2 + \ldots+ N_n~.
\eea
This theory can be realised as the worldvolume of the D3-branes in the following configuration \cite{Hanany:1996ie}:
\bea 
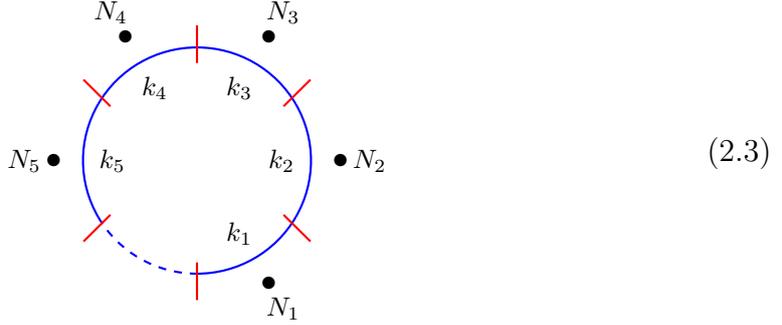
\label{braneKN}
\begin{tikzpicture}[baseline, align=center,node distance=0.5cm]
\def \n {6}
\def \radius {1.5cm}
\def \margin {0} 
\foreach \s in {1,...,5}
{
  \node[draw=none] (\s) at ({360/\n * (\s - 2)+30}:{\radius-10}) {};
  \node[draw=none]  at ({360/\n * (\s - 2)}:{\radius-11}) {{\footnotesize $k_{\s}$}};
  \node[draw=none]  at ({360/\n * (\s - 2)}:{\radius+11}) {$\bullet$};
  \node[draw=none]  at ({360/\n * (\s - 2)}:{\radius+22}) {{\footnotesize $N_{\s}$}};
  \draw[-, >=latex,blue,thick] ({360/\n * (\s - 3)+\margin+30}:\radius) 
    arc ({360/\n * (\s - 3)+\margin+30}:{360/\n * (\s-2)-\margin+30}:\radius);
}
\node[draw=none, circle] (last) at ({360/3 * (3 - 1)+30}:{\radius-10}) {};
\draw[dashed, >=latex,thick,blue] ({360/6 * (5 -2)+\margin+30}:\radius) 
    arc ({360/6 * (5 -2)+\margin+30}:{360/6 * (5-1)-\margin+30}:\radius);
    \node[draw=none,  below right= of 1] (f1) {};
\node[draw=none, above right= of 2] (f2) {};
\node[draw=none, above = of 3] (f3) {};
\node[draw=none, above left= of 4] (f4) {};
\node[draw=none,  below left= of 5] (f5) {};
\node[draw=none,  below = of last] (f6) {};
\draw[-, >=latex,red, thick] (1) to (f1);
\draw[-, >=latex,red, thick] (2) to (f2);
\draw[-, >=latex,red, thick] (3) to (f3);
\draw[-, >=latex,red, thick] (4) to (f4);
\draw[-, >=latex,red, thick] (5) to (f5);
\draw[-, >=latex,red, thick] (last) to (f6);
\end{tikzpicture}
\eea
where each blue line denotes the D3-branes, each red line denotes an NS5-branes, and each black dot with the label $N_i$ denotes the $N_i$ D5-branes.  The branes span the following directions:
\bea
\begin{array}{c|cccccccccc}
\hline
                     & 0 & 1 & 2 & 3 & 4 & 5 & 6 & 7 & 8 & 9 \\
\hline                     
\mathrm{D3} & \mathrm{X} & \mathrm{X} & \mathrm{X} &~&~ &~ &\mathrm{X} &~ &~ &~  \\
\mathrm{NS5}&\mathrm{X}& \mathrm{X} & \mathrm{X} &~ &~ &~ &~ &\mathrm{X} &\mathrm{X}& \mathrm{X}  \\
\mathrm{D5} &\mathrm{X}& \mathrm{X} & \mathrm{X} &\mathrm{X} &\mathrm{X} &\mathrm{X} &~ &~ &~& ~\\
\hline
\end{array}
\eea
where the $6$-th direction corresponds to the circular direction.

From quiver \eref{KN}, we can read off the information about the gauge field at infinity $U_\infty$ and the gauge field at the origin $U_0$ as follows \cite{Witten:2009xu}.  The number of eigenvalues of $U_\infty$ that equal to $e^{2\pi i \ell/n}$ (for $\ell =1, \ldots, n$) is $N_\ell$.  Here $N_\ell$ also has an interpretation as the number of D5-branes with linking number $\ell$.  The number of eigenvalues of $U_0$ that equal to $e^{2\pi i \ell/n}$ is 
\bea
\beta_\ell &= N_\ell + k_{\ell+1} +k_{\ell-1} -2k_{\ell}~, \qquad \ell = 1,\ldots, n~.
\eea
Note that $\beta_\ell$ is the difference between the linking numbers of the $(\ell+1)$-th and the $\ell$-th NS5-branes.  From now on and in the main text, we refer to the paritition $(N_1, \ldots, N_n)$ of $N$ as the {\it framing} of the $PU(N)$ instantons on $\BC^2/\BZ_n$.  Indeed the framing is cyclic.

For simplicity of the discussion, throughout section \ref{sec:singular} we take
\bea
k_1 = k_2 = \ldots = k_n = k~.
\eea
and use the terminology that $k$ is the instanton number.  In this case, the monodromies $U_0$ and $U_\infty$ have the same eigenvalues.  As a result, it is sufficient to specify the instanton configuration just by the framing $(N_1, \ldots, N_n)$.  In this case, the monodromy breaks $PU(N) \cong U(N)/U(1)$ into the residual symmetry $(U(N_1) \times U(N_2) \times \dots \times U(N_n))/U(1)$.

\subsection*{The Coulomb branch of the Kronheimer-Nakajima quiver}

So far we have discussed about the Higgs branch of the KN quiver. Let us now study the Coulomb branch of such a quiver. 
A simple way to do so is to apply three dimensional mirror symmetry \cite{Intriligator:2013lca} to quiver \eref{KN} and study the Higgs branch of the resulting quiver.  The former amounts to apply an S-duality on the configuration \eref{braneKN}, under which the NS5-branes and the D5-branes are exchanged \cite{Hanany:1996ie}.

The result is a circular quiver with $N$ $U(k)$ gauge groups with one flavour of fundamental hypermultiplet under each of the $N_1$-th, $N_2$-th,  $\ldots$, $N_n$-th gauge groups.  By mirror symmetry, the Coulomb branch of the KN quiver \eref{KN} is therefore the Higgs branch of this resulting quiver and it describes
\bea \label{CoulombKNquivconf}
&\text{the moduli space of $k$ $PU(n)$ instantons on $\BC^2/\BZ_N$}  \nn \\
& \text{with framing $(0^{N_1-1},1, 0^{N_2-1},1,\ldots, 0^{N_n-1},1)$.}
\eea
Comparing \eref{KN} and \eref{CoulombKNquivconf}, one observes that the roles of the gauge group and the orbifold type get exchanged  \cite{deBoer:1996mp, Porrati:1996xi, Hanany:1999sj, Dey:2013nf} under mirror symmetry.

The framing in \eref{CoulombKNquivconf} determines the residual symmetry of $PU(n)$.  If $N_i >0$ for all $i =1, 2, \ldots, n$, then $PU(n)$ is broken to its maximal abelian subgroup $U(1)^{n-1}$.  On the other hand, if some $N_i$ vanishes, $PU(n)$ is broken to a subgroup that contains a non-abelian symmetry, \eg~ for $n=3$, if $N_1=2$, $N_2=0$ and $N_3= 1$, then the framing is $(0,2,1)$ and $PU(3) \cong U(3)/U(1)$ is thus broken to $(U(2) \times U(1))/U(1)$ .

There are some interesting special cases to consider:
\bi
\item If one of the $N_i$'s is equal to one and the other $N_i$'s are zero, the Coulomb branch of  can be identified with the moduli space of $k$ $SU(n)$ instantons on $\BC^2$; in agreement with \cite{Cremonesi:2014xha}.
\item If one of the $N_i$'s is equal to $N$ and the other $N_i$'s are zero, the symmetry of the Coulomb branch is $U(1) \times SU(n)$ for $N \geq 3$ and $SU(2) \times SU(n)$ for $N=1,2$, where the $U(1)$ and $SU(2)$ are associated with the isometry of $\BC^2/\BZ_n$.

If in addition we set $k=1$, the Coulomb branch of \eref{KN} is isomorphic to 
\bea \label{1pureSUN}\BC^2/\BZ_N \times \CN_{SU(n)}~, \eea 
where $\CN_{SU(n)}$ is the reduced moduli space of one $SU(n)$ instanton on $\BC^2$, which is is the minimal nilpotent orbit of $SU(n)$ \cite{kronheimer1990, Br, KS1, Gaiotto:2008nz}.  On the other hand, the Higgs branch of \eref{KN} in this case is 
\bea \label{Higgs1pureSUN} \BC^2/\BZ_n \times \CN_{SU(N)}~, \eea 
The feature \eref{Higgs1pureSUN} of the moduli space was in fact pointed out in (2.69) of \cite{Dey:2013fea}.
\ei

\subsection{A simple group $G$ and $\Gamma =\BZ_n$ or $G=PU(N)$ and $\Gamma= \widehat{A}_n, \; \widehat{D}_n, \;\widehat{E}_{6,7,8}$} \label{gensimGC2Zn}
This result can in fact be generalised to the moduli space of a simple gauge group $G$ instantons on $\BC^2/\BZ_n$ and that of $PU(n)$ instantons on $\BC^2/\Gamma$, with $\Gamma= \widehat{D}_n, \; \widehat{E}_{6}, \; \widehat{E}_7,\; \widehat{E}_8$.  In each case, we specify the monodromy of the gauge field at infinity by stating the residual symmetry of the instanton gauge group.

As a natural generalisation of the KN quiver, we focus on the following quiver diagram:
\begin{quote}
The affine Dynkin diagram of the group $G$ with the gauge groups $U(k \mathfrak{a}^\vee_i)$, where $\mathfrak{a}^\vee_i$ ($i=0,1,\ldots, {\rm rank}\; G$) is the dual Coxeter label of the $i$-th node, and with $n_i$ flavours of fundamental hypermultiplets attached to the $i$-th node such that
\bea \label{genDyn}
n = \sum_{i=0}^{{\rm rank}\; G} \mathfrak{a}_i n_i~,
\eea
where $\mathfrak{a}_i$ are the Coxeter label of the $i$-th node.\footnote{The information about ordering of the nodes, coxeter labels and dual coxeter labels for the affine Dynkin diagrams of simple groups $G$ can be found in Figure 14.1 on Page 563 of \cite{philippe1997conformal}.}
\end{quote}

For $G$ a simply laced group of types $A$, $D$ or $E$, this is a conventional quiver diagram possessing a known Lagrangian description, whose moduli space was studied in various papers, \eg~ \cite{Intriligator:1996ex, Porrati:1996xi, deBoer:1996mp, Nakajima:2015txa}.   
On the other hand, if $G$ is a non-simply laced group of type $B$, $C$, $F$ or $G$, the Lagrangian description of this quiver is not currently known, due to the presence of multiple-laces in the Dynkin diagram, and we refer to such a diagram as a generalised quiver diagram.  As was discussed in \cite{Cremonesi:2013lqa, Cremonesi:2014vla, Cremonesi:2014xha}, the Coulomb branch of such quiver diagrams has been proven to be useful in the study of the moduli space of instantons on flat space.  In the following, we generalise the previous results to the moduli space of instantons on a singular orbifold.

Given quiver diagram \eref{genDyn}, we state the following results regarding its Coulomb and Higgs branch:
\ben
\item  The  {\it Coulomb branch} of the above quiver describes the moduli space of $k$ $G$ instantons on $\BC^2/\BZ_n$ with monodromy at infinity such that the symmetry $G$ is broken to a subgroup $H$ of $G$, where 
\begin{quote} 
the non-abelian factors of $H$ are constructed by removing the simple roots ${\vec \alpha}_i$ associated with the nodes where $n_i \neq 0$ and the abelian factors of $H$ are present such that the total rank of $H$ is equal to the rank of $G$.
\end{quote}
For $n=1, 2$, the isometry of the Coulomb branch is $SU(2) \times H$, where the $SU(2)$ factor corresponds to the isometry of $\BC^2$ or $\BC^2/\BZ_2$.  For $n \geq 3$ the isometry of the Coulomb branch is $U(1) \times H$, where the $U(1)$ factor corresponds to the isometry of $\BC^2/\BZ_n$ with $n \geq 3$.
\item For $G=A, D, E$, the {\it Higgs branch} of the above quiver describes the moduli space of $PU(n)$ instantons on $\BC^2/\widehat{G}$, where $\widehat{G}$ is the discrete group that is in McKay correspondence to $G$,  with monodromy at infinity such that the group $PU(n)$ is broken to its subgroup determined by the flavour nodes in the quiver diagram.  
\een
Observe that the instanton gauge group and the orbifold type get exchanged as one goes from the Higgs branch to the Coulomb branch and vice versa \cite{deBoer:1996mp, Porrati:1996xi, Hanany:1999sj, Dey:2013nf}.

Many of these statements can be conveniently checked using Hilbert series.  For a simply laced group $G$, the Hilbert series of the Coulomb branch mentioned in item 1 can be computed using the method described in \cite{Cremonesi:2013lqa}.  For a non-simply laced group $G$, the method of \cite{Cremonesi:2014xha} can be applied.  We review such methods in Appendix \ref{app:Coulomb} and provide several explicit examples in section \ref{sec:examples}.

The quaternionic dimension of the Coulomb branch of quiver \eref{genDyn} is
\bea
{\rm dim}_{\BH} ~\text{Coulomb of \eref{genDyn}} = k \sum_{i=0}^{{\rm rank}\; G} a^\vee_i = k h^\vee_G~,
\eea
where $h^\vee_G$ is the dual coxeter number of $G$.  This is to be expected as the dimension of the moduli space of $k$ $G$ instantons on $\BC^2/\BZ_n$.  On the other hand, for $G=A, D, E$, the quaternionic dimension of the Higgs branch of quiver \eref{genDyn} is
\bea
{\rm dim}_{\BH} ~\text{Higgs of \eref{genDyn}} =  k \sum_{i=0}^{{\rm rank}\; G} a^\vee_i n_i =  k \sum_{i=0}^{{\rm rank}\; G} a_i n_i   = kn =k h^\vee_{SU(n)},
\eea
which is to be expected as the dimension of the moduli space of $k$ $PU(n)$ instantons on $\BC^2/\widehat{G}$.

In certain special cases, the moduli space of quiver \eref{genDyn} possesses the following important features:
\bi
\item If $n_0=1$ and $n_i=0$ for $i \neq 0$, the Coulomb branch of  can be identified with the moduli space of $k$ $G$ instantons on $\BC^2$.  This is discussed in detail in \cite{Cremonesi:2014xha}.  

If in addition we take $k=1$, the Coulomb branch is isomorphic to $\BC^2 \times \CN_G$, where $\CN_G$ is  the minimal nilpotent orbit of $G$ \cite{kronheimer1990, Br, KS1, Gaiotto:2008nz}, and the Higgs branch for $G=A, D, E$ is isomorphic to $\BC^2/\widehat{G}$, where $\widehat{G}$ is the discrete group that is in McKay correspondence to $G$.
\item If $n_0=n$ and $n_i=0$ for $i \neq 0$, the symmetry of the Coulomb branch is $U(1) \times G$ for $n \geq 3$ and $SU(2) \times G$ for $n=1,2$.  

If in addition we set $k=1$, the Coulomb branch of \eref{genDyn} is isomorphic to 
\bea \BC^2/\BZ_n \times \CN_{G} \label{pure1G} \eea
and, for $G=A, D, E$, the Higgs branch of \eref{genDyn} is isomorphic to 
\bea \BC^2/\widehat{G} \times \CN_{SU(n)}~. \label{Higgspure1G} \eea
\ei

In the following, we present some examples and compute the Hilbert series to demonstrate the above general rule.  
\subsection{Examples} \label{sec:examples}
Below we provide three examples involving $D_4$, $B_3$ and $G_2$ affine Dynkin diagrams.  In the first example, the moduli space of $SO(8)$ instantons on $\BC^2/\BZ_n$ and that of $PU(N)$ instantons on $\BC^2/\hat{D}_4$ are discussed.  In the second and the third examples, we demonstrate the use of generalised quiver diagrams, which involving non-simple laces, in the study of moduli spaces of $SO(7)$ and $G_2$ instantons on $\BC^2/\BZ_n$.

\subsubsection{Flavoured $D_4$ affine Dynkin diagrams}  
Below we present the quivers diagrams whose {\it Coulomb branches} correspond to the moduli space correspond to $k$ $SO(8)$ instantons on $\BC^2/\BZ_2$ with various monodromies at infinity.  The symmetry of the Coulomb branch is indicated below each quiver diagram.   The $SO(8)$ symmetry is broken to $SO(8)$, $SO(2)\times SO(6)$ and $SO(4) \times SO(4)$, respectively from left to right.  The factor of $SU(2)$ that appears under each diagram corresponds to the isometry of $\BC^2/\BZ_2$. These symmetries can also be read off from the Coulomb branch Hilbert series\footnote{Some of which were computed in \cite{Dey:2014tka}.}, whose computations are presented below.  They are in accordance with the general rule stated earlier.

\bea \label{quivD4s}
\begin{array}{ccc}
\begin{tikzpicture}[align=center,node distance=0.5cm]
  \node[draw, circle] (4) at (0,0) {{\footnotesize $2k$}};
  \node[draw, circle] (0) at (-1,1) {{\footnotesize $k$}};
  \node[draw, circle] (1) at (-1,-1) {{\footnotesize $k$}};
  \node[draw, circle] (2) at (1,-1) {{\footnotesize $k$}};
  \node[draw, circle] (3) at (1,1) {{\footnotesize $k$}};
  \node[draw, rectangle, left= of 0] (flv) {{\footnotesize $2$}};
\foreach \s in {0,...,3}
{
  \draw[-] (4) to (\s);
}
\draw[-] (flv) to (0);
\node[draw=none] at (0,-2) {{\footnotesize $SU(2) \times SO(8)$}};
\end{tikzpicture} 
& \qquad  \quad
\begin{tikzpicture}[align=center,node distance=0.5cm]
  \node[draw, circle] (4) at (0,0) {{\footnotesize $2k$}};
  \node[draw, circle] (0) at (-1,1) {{\footnotesize $ k$}};
  \node[draw, circle] (1) at (-1,-1) {{\footnotesize $ k$}};
  \node[draw, circle] (2) at (1,-1) {{\footnotesize $k$}};
  \node[draw, circle] (3) at (1,1) {{\footnotesize $k$}};
  \node[draw, rectangle, left= of 0] (flv1) {{\footnotesize $1$}};
  \node[draw, rectangle, left= of 1] (flv2) {{\footnotesize $1$}};
\foreach \s in {0,...,3}
{
  \draw[-] (4) to (\s);
}
\draw[-] (flv1) to (0);
\draw[-] (flv2) to (1);
\node[draw=none] at (0,-2) {{\footnotesize $SU(2) \times (SO(2) \times SO(6))$}};
\end{tikzpicture}
& \qquad 
\begin{tikzpicture}[align=center,node distance=0.5cm]
  \node[draw, circle] (4) at (0,0) {{\footnotesize $2k $}};
  \node[draw, circle] (0) at (-1,1) {{\footnotesize $k$}};
  \node[draw, circle] (1) at (-1,-1) {{\footnotesize $k$}};
  \node[draw, circle] (2) at (1,-1) {{\footnotesize $k$}};
  \node[draw, circle] (3) at (1,1) {{\footnotesize $k$}};
  \node[draw, rectangle, below= of 4] (flv1) {{\footnotesize $1$}};
\foreach \s in {0,...,3}
{
  \draw[-] (4) to (\s);
}
\draw[-] (flv1) to (4);
\node[draw=none] at (0,-2) {{\footnotesize $SU(2) \times (SO(4) \times SO(4))$}};
\end{tikzpicture}
\end{array}
\eea
The rightmost diagram of \eref{quivD4s} can be redrawn as
\bea \label{fivefinger}
\begin{tikzpicture}[align=center,node distance=0.5cm]
  \node[draw, circle] (4) at (0,0) {{\footnotesize $2k $}};
  \node[draw, circle] (0) at (-1,1) {{\footnotesize $ k$}};
  \node[draw, circle] (1) at (-1,-1) {{\footnotesize $k$}};
  \node[draw, circle] (2) at (1,-1) {{\footnotesize $k$}};
  \node[draw, circle] (3) at (1,1) {{\footnotesize $k$}};
  \node[draw, circle, below= of 4] (flv1) {{\footnotesize $1$}};
\foreach \s in {0,...,3}
{
  \draw[-] (4) to (\s);
}
\draw[-] (flv1) to (4);
\end{tikzpicture}
\eea
with the overall $U(1)$ decoupled (for example, from the middle $U(2k)$ gauge node).   It should be noted that this theory is actually a $3d$ mirror theory \cite{Benini:2010uu} of the $3d$ Sicilian theory (the class $\CS$ theory compactified on $S^1$) of type $A_{2k-1}$ with punctures:
\bea
(k,k), \; (k,k), \; (k,k), \; (k,k), \; (2k-1,1)~.
\eea

\subsubsection*{The Coulomb branch Hilbert series for $k=1$}
Let us compute the Coulomb branch Hilbert series of \eref{quivD4s} in the case of $k=1$.  For the sake of brevity in writing the formulae below, let us define
\bea
&H[D_4](\Delta_{m}) \nn \\
&=\sum_{u_0 \in \BZ} \;  \sum_{u_1 \in \BZ}\;  \sum_{u_3 \in \BZ}\;  \sum_{u_4 \in \BZ} \; \sum_{n_1 \geq n_2 > -\infty} t^{\Delta_{m} -2|n_1-n_2|+\sum_{i=1}^2 (|u_0-n_i|+|u_1-n_i| + |u_3-n_i|+|u_4-n_i|)} \nn \\
& \qquad z_0^{u_0} z_1^{u_1} z_2^{n_1+n_2} z_3^{u_3} z_4^{u_4} \times P_{U(2)}(t; n_1, n_2) (1-t^2)^{-4}~,
\eea
where $\Delta_m$ denotes the contribution from the fundamental matter attached to the $D_4$ affine quiver.  It may depend on $u$'s and $n$'s.

The Coulomb branch Hilbert series of the left quiver of \eref{quivD4s} is
\bea \label{C2Z2timesnilso8}
&H^{\text{Coul}}_{\eref{quivD4s}, \text{left}} (t; z_0, \ldots, z_4) \nn \\
&= H[D_4](2|u_0|) \nn \\
&= \left( \sum_{p_1=0}^\infty \chi^{SU(2)}_{[2p_1]}(x) t^{2p_1} \right)\left( \sum_{p_2=0}^\infty \chi^{SO(8)}_{[0,p_2,0,0]}(y_1, \ldots, y_4) t^{2p_2} \right)~,
\eea
where
\bea
x^2= z_0 z_1 z_2^2 z_3 z_4~, \qquad y_1= z_1,  \qquad y_2= z_2, \qquad y_3 = z_3, \qquad y_4=z_4~.
\eea
This is the Hilbert series of $(\BC^2/\BZ_2) \times \CN_{SO(8)}$, where $\CN_{SO(8)}$ is  the minimal nilpotent orbit of $SO(8)$.  

The Coulomb branch Hilbert series of the middle quiver of \eref{quivD4s} is
\bea
&H^{\text{Coul}}_{\eref{quivD4s}, \text{mid}} (t; z_0, \ldots, z_4) \nn \\
&= H[D_4](|u_0|+|u_1|) \nn \\
&= \PE \Big[ \Big(\chi^{SU(2)}_{[2]}(x)+\chi^{SO(6)}_{[0,1,1]}(\vec y) + 1\Big)t^2 +(\chi^{SU(2)}_{[1]} \chi^{SO(2)}_{[1]}(w) \chi^{SO(6)}_{[1,0,0]} (\vec y) ) t^3 \nn \\
& \qquad \qquad -(\chi^{SO(6)}_{[0,1,1]}(\vec y) + 2)t^4+ \ldots \Big]
\eea
where
\bea
x^2= z_0 z_1 z_2^2 z_3 z_4~, \qquad w_1^2= z_0 z_1^{-1}~, \qquad y_1 =z_2~, \qquad y_2 = z_3~, \qquad y_3 =z_4~.
\eea
This is in agreement with (4.53) of \cite{Dey:2013fea}.

The Coulomb branch Hilbert series of the right quiver of \eref{quivD4s} is
\bea
&H^{\text{Coul}}_{\eref{quivD4s}, \text{right}} (t; z_0, \ldots, z_4) \nn \\
&= H[D_4](|n_1|+|n_2|) \nn \\
&= \PE \Big[ \Big( \chi^{SU(2)}_{[2]}(x) +\chi^{SO(4)}_{[2,0]+[0,2]}(\vec w) +\chi^{SO(4)}_{[2,0]+[0,2]}(\vec y) \Big) t^2   \nn \\
& \qquad \qquad +  \chi^{SU(2)}_{[1]}(x)\chi^{SO(4)}_{[1,1]}(\vec w) \chi^{SO(4)}_{[1,1]}(\vec y) t^3-4t^4 +\ldots   \Big]~,
\eea
where 
\bea
x^2= z_0 z_1 z_2^2 z_3 z_4~, \qquad y_1^2= z_0,  \qquad y_2^2= z_1, \qquad w_1^2 = z_3, \qquad w_2^2=z_4~.
\eea
If we set $z_0 = \ldots= z_4 =1$, we obtain the unrefined Hilbert series
\bea
&H^{\text{Coul}}_{\eref{quivD4s}, \text{right}} (t; z_0=1, \ldots, z_4=1) \nn \\
& = 1 + 15 t^2 + 32 t^3 + 116 t^4 + 352 t^5 + 863 t^6 + \ldots~.
\eea

\subsubsection*{An example for $SO(8)$ instantons on $\BC^2/\BZ_3$}
In addition to the quivers depicted in \eref{quivD4s}, let us consider the following diagram
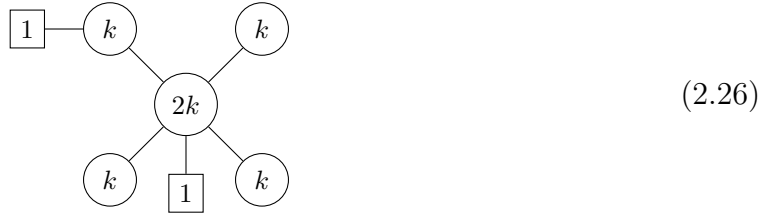
\bea \label{SO8C2Z3}
\begin{tikzpicture}[baseline, align=center,node distance=0.5cm]
  \node[draw, circle] (4) at (0,0) {{\footnotesize $2k $}};
  \node[draw, circle] (0) at (-1,1) {{\footnotesize $k$}};
  \node[draw, circle] (1) at (-1,-1) {{\footnotesize $k$}};
  \node[draw, circle] (2) at (1,-1) {{\footnotesize $k$}};
  \node[draw, circle] (3) at (1,1) {{\footnotesize $k$}};
  \node[draw, rectangle, below= of 4] (flv1) {{\footnotesize $1$}};
    \node[draw, rectangle, left= of 0] (flv2) {{\footnotesize $1$}};
\foreach \s in {0,...,3}
{
  \draw[-] (4) to (\s);
}
\draw[-] (flv1) to (4);
\draw[-] (flv2) to (0);
\node[draw=none] at (0,-2) {{\footnotesize $U(1) \times (U(1) \times SU(2) \times SO(4) )$}};
\end{tikzpicture}
\eea
The Coulomb branch of this quiver describes the moduli space of $k$ $SO(8)$ instantons on $\BC^2/\BZ_3$ with monodromy at infinity such that $SO(8)$ is broken to $U(1) \times SU(2) \times SO(4)$.  According to the general rule stated above, the $SU(2)\times SO(4)$ factor follows from the removal of the simple roots associated with the top left and the middle nodes and the $U(1)$ factor is precisely the abelian factor whose existence makes the rank add up to $4$.  The symmetry of the Coulomb branch is $U(1) \times (U(1) \times SU(2) \times SO(4) )$, where the first $U(1)$ is associated with the isometry of $\BC^2/\BZ_3$.  This symmetry can be confirmed using the Coulomb branch Hilbert series.  For example, for $k=1$, this is given by
\bea
H^{\text{Coul}}_{\eref{SO8C2Z3}}(t; z_0, \ldots, z_4)=H[D_4](|u_0|+|u_2|)~.
\eea
For brevity, we present the Hilbert series when $z_i$ are set to $1$ for all $i=0,\ldots, 4$:
\bea
H^{\text{Coul}}_{\eref{SO8C2Z3}}(t; \{z_i = 1 \})=1+11 t^2+20 t^3+82 t^4+\ldots~. \label{CoulquivSO8C2Z3}
\eea
Note that the coefficient $11$ of $t^2$ is indeed the sum of the dimensions of the adjoint representations of each factor in $U(1) \times (U(1) \times SU(2) \times SO(4) )$, \ie~ $1+1+3+6$.

\subsubsection*{Mirror theories of \eref{quivD4s} and \eref{SO8C2Z3}}
The above instanton moduli spaces of instantons can be realised from the {\it Higgs branch} of the following quiver diagram (see section 4.2 of \cite{Intriligator:1997kq} and section 4.1 of \cite{Dey:2013fea}):
\bea \label{SOUSpUSpSO}
\begin{tikzpicture}[align=center, node distance=0.8cm,minimum size=1.2cm]
  \node[draw, rectangle] (1) at (0,0) {{\footnotesize $SO(p)$}};
  \node[draw, circle, right=of 1] (2) {{\scriptsize $USp(2k)$}};
  \node[draw, circle, right=of 2] (3) {{\scriptsize $USp(2k)$}};
  \node[draw, rectangle, right= of 3] (4) {{\footnotesize $SO(8-p)$}};
  \draw[-] (1) to (2);
  \draw[-] (2) to (3);
  \draw[-] (3) to (4);
\end{tikzpicture} 
\eea
where $p=0,2,4$.  For $p=2,4$, this quiver is indeed a three-dimensional mirror theory for the quivers depicted in \eref{quivD4s}, respectively from left to right.  Note that the symmetry of the Higgs branch of this quiver is indeed $SU(2) \times SO(p) \times SO(8-p)$.  We check that the Coulomb branch Hilbert series computed from \eref{quivD4s} are in agreement with those computed from the Higgs branch of \eref{SOUSpUSpSO}.  

The case of $p=0$ of the deserves a special discussion\footnote{This paragraph is added in version 3 of this article on the {\tt arXiv}. I thank Amihay Hanany for pointing this out.}.  In this case, the left $USp(2k)$ gauge node has $2k$ flavours of hypermultiplets transforming under its fundamental representation and so the Higgs branch of \eref{SOUSpUSpSO} is expected to be the union of two cones \cite{Ferlito:2016grh}. For simplicity, let us discuss explicitly both cones in the case of $k=1$.  (The discussion can be easily generalised to the higher values of $k$.) One of the cone can be identified with the product of $\BC^2/\BZ_2$ and (the closure of) the minimal nilpotent orbit of $SO(8)$; this is identified with the Coulomb branch of the leftmost quiver of \eref{quivD4s}.  The other cone turns out to be the next-to-minimal nilpotent orbit of $SO(8)$; this is identified with the Coulomb branch of the quiver depicted in \cite[Figure 9]{Hanany:2018xth}\footnote{In a contemporary terminology, this is also known as a {\it magnetic quiver} \cite{Cremonesi:2015lsa, Cabrera:2018jxt, Hanany:2018uhm, Cabrera:2019izd} associated with the next-to-minimal nilpotent orbit of $SO(8)$.}:  
\be\label{magquiv}
\begin{tikzpicture}[baseline, align=center,node distance=0.5cm]
  \node[draw, circle] (3) at (0,0) {{\footnotesize $2$}};
  \node[draw, circle] (0) at (-1,0) {{\footnotesize $2$}};
  \node[draw, circle] (1) at (1,-1) {{\footnotesize $1$}};
  \node[draw, circle] (2) at (1,1) {{\footnotesize $1$}};
  \node[draw, rectangle, left= of 0] (flv) {{\footnotesize $2$}};
\foreach \s in {0,...,3}
{
  \draw[-] (3) to (\s);
}
\draw[-] (flv) to (0);
\end{tikzpicture} 
\ee
To see the Hilbert series of each cone, we recall the Hilbert series of the $USp(2)$ gauge theory with $2$ flavours \cite[(2.5)]{Ferlito:2016grh}:
\be
\PE[ \chi^{SU(2)}_{[2]} (u) t^2 -t^4]+ \PE[ \chi^{SU(2)}_{[2]} (v) t^2 -t^4] -1~.
\ee
where the first two terms are the contributions of the two cones and $-1$ takes into account of the intersection.
To obtain \eref{SOUSpUSpSO} with $p=0$ and $k=1$, we gauge the $USp(2)$ subgroup of the flavour symmetry of the said $USp(2)$ gauge theory and couple it to $4$ flavours of hypermultiplets in the fundamental representation.  The Hilbert series of the first cone of \eref{SOUSpUSpSO} with $p=0$ and $k=1$ is
\bea
& \oint_{|v|=1} \frac{dv}{2 \pi v} (1-v^2) \PE[ \chi^{SU(2)}_{[2]} (u) t^2 -t^4] \times\nn \\
&\qquad\qquad  \PE[\chi^{SU(2)}_{[1]} (v) \chi^{SO(8)}_{[1,0,0,0]} (\vec f) t - \chi^{SU(2)}_{[2]} (v) t^2 ] \nn\\
& =\PE[ \chi^{SU(2)}_{[2]} (u) t^2 -t^4]  \times \sum_{p=0}^\infty \chi^{SO(8)}_{[0,p,0,0]} (\vec f) t^{2p}~, 
\eea
where this is indeed the Hilbert series of the product of $\BC^2/\BZ_2$ and the minimal nilpotent orbit of $SO(8)$.  The Hilbert series of the other cone of \eref{SOUSpUSpSO} with $p=0$ and $k=1$ is
\bea
& \oint_{|v|=1} \frac{dv}{2 \pi v} (1-v^2)  \PE[ \chi^{SU(2)}_{[2]} (v) t^2 -t^4]  \times\nn \\
&\qquad\qquad \PE[\chi^{SU(2)}_{[1]} (v) \chi^{SO(8)}_{[1,0,0,0]} (\vec f) t - \chi^{SU(2)}_{[2]} (v) t^2 ]~.
\eea
As a result of the integration, we obtain the Hilbert series of the next-to-minimal nilpotent orbit of $SO(8)$, whose highest weight generating function and the unrefined Hilbert series (\ie~ all components of $\vec f$ are set to 1) are given by \cite[Row 4, Table 15]{Hanany:2016gbz}, namely
\bea
\frac{1}{(1-m_1^2 t^4)(1-m_2 t^2)}~, \quad \frac{(1 + t^2)^2 (1 + 14 t^2 + 36 t^4 + 14 t^6 + t^8)}{(1-t^2)^{12}}~.
\eea
It can also indeed be checked that this Hilbert series is also equal to the Coulomb branch Hilbert series of \eref{magquiv}.

Note that for the middle quiver of \eref{quivD4s} has two mirror theories (see sections 4.1 and 4.2 of \cite{Intriligator:1997kq}, section 4.2 of \cite{Dey:2013fea} and Fig. 22 on Page 47 of \cite{Dey:2014tka}):
\bea
&\begin{tikzpicture}[baseline, align=center, node distance=0.8cm,minimum size=1.2cm]
  \node[draw, rectangle] (1) at (0,0) {{\footnotesize $SO(2)$}};
  \node[draw, circle, right=of 1] (2) {{\scriptsize $USp(2k)$}};
  \node[draw, circle, right=of 2] (3) {{\scriptsize $USp(2k)$}};
  \node[draw, rectangle, right= of 3] (4) {{\footnotesize $SO(6)$}};
  \draw[-] (1) to (2);
  \draw[-] (2) to (3);
  \draw[-] (3) to (4);
\end{tikzpicture} 
\\
& \hspace{2.5cm} \begin{tikzpicture}[baseline, align=center, node distance=0.8cm,minimum size=1.2cm]
  \node[draw, rectangle] (1) at (0,0) {{\scriptsize $U(4)$}};
  \node[draw, circle, right=of 1] (2) {{\scriptsize $U(2k)$}};
  \node[draw=none] at (3.4,-0.5) {{\scriptsize $A_2$}};
  \node[draw=none] at (3.4,0.5) {{\scriptsize $A_1$}};
  \draw[-] (1) to (2);
  \draw [-] (2) edge [out=270,in=0,loop,looseness=4] (2);
  \draw [-] (2) edge [out=0,in=90,loop,looseness=4] (2);
\end{tikzpicture} 
\eea
where $A_1$ and $A_2$ denote antisymmetric hypermultiplet under gauge group $U(2)$.  The Higgs branch of either theory describes $k$ $SO(8)$ instantons on $\BC^2/\BZ_2$ with $SO(8)$ broken to $SO(2) \times SO(6)$, whose algebra is isomorphic to $U(4)$.

A mirror theory of \eref{SO8C2Z3} is depicted below.
\vspace{-0.7cm}
{\small
\bea \label{mirrorSO8C2Z3}
\begin{tikzpicture}[baseline, align=center, node distance=0.9cm,minimum size=1.2cm]
  \node[draw, circle] (1) {{\scriptsize $U(2k)$}};
  \node[draw, circle, right=of 1] (2) {{\tiny $USp(2k)$}};
  \node[draw, rectangle, left=of 1] (3){{\footnotesize $U(2)$}};
  \node[draw, rectangle, right=of 2] (4) {{\footnotesize $SO(4)$}};
  \node[draw=none] at (0,1.2) {{\scriptsize $A$}};
  \draw [-] (1) edge [out=135,in=45,loop,looseness=3] (1);
  \draw[-] (1) to (2);
  \draw[-] (1) to (3);
  \draw[-] (2) to (4);
\end{tikzpicture} 
\eea}
where $A$ denotes the rank two antisymmetric hypermultiplet under the $U(2k)$ gauge group.  Indeed, the symmetry of the Higgs branch is $U(1) \times U(2) \times SO(4)$, where $U(1)$ is associated with the antisymmetric hypermultiplet.  This symmetry agrees with the Coulomb branch symmetry of quiver \eref{SO8C2Z3}.\footnote{Indeed one can also check that the Higgs branch Hilbert series of \eref{mirrorSO8C2Z3} agrees with \eref{CoulquivSO8C2Z3}.}

\subsubsection*{The Higgs branch of \eref{quivD4s}}
The {\it Higgs branch} of the quivers in \eref{quivD4s} corresponds to the moduli space of $k$ $PU(2)$ instantons on $\BC^2/\widehat{D}_4$ with monodromy at infinity such that $PU(2) \cong U(2)/U(1)$ symmetry is broken to $PU(2)$, $(U(1) \times U(1))/U(1)$ and empty, respectively from left to right.\footnote{These moduli spaces can also be realised from the Coulomb branch of \eref{SOUSpUSpSO} with $p=0,2,4$ respectively.}  These symmetries are manifest from the quiver diagrams\footnote{For the rightmost quiver in \eref{quivD4s}, one can see this from quiver \eref{fivefinger}.}.  In the following, we examine the Higgs branches of the leftmost and the rightmost quivers in detail. \\~\\
\noindent{\it The leftmost quiver of \eref{quivD4s}.} For $k=1$, the Higgs branch Hilbert series can be computed as follows:
\bea
&H^{\text{Higgs}}_{\eref{quivD4s}, \text{left}} (t, y)  \nn \\
&= \oint_{|z|=1} \frac{dz}{2 \pi i z} (1-z^2)  \oint_{|q|=1}  \frac{dq}{2 \pi i q} \oint_{|w|=1}  \frac{dw}{2 \pi i w} \times  \\
& \qquad \frac{\PE[ \chi^{SU(2)}_{[2]} (z) t^2 -t^4 ]^3 \PE[(q w^{-1} +q^{-1} w)\chi^{SU(2)}_{[1]}(z) t ] \PE[(y+y^{-1})(w+w^{-1})t]}{\PE[(z^2+2+z^{-2})t^2]\PE[t^2] }~,\nn
\eea
where $(z, q)$ are $U(2) \cong SU(2) \times U(1)$ gauge fugacities associated with the middle node; $w$ is the $U(1)$ gauge fugacity associated with top left node; and $y$ is the fugacity of the $SU(2)$ flavour symmetry.  The first factor in the numerator denotes three copies of the Higgs branch of $U(1)$ gauge theory with 2 flavours (\ie~ three copies of $\BC^2/\BZ_2$).  The second and the third factors in the numerator denote the contribution from the remaining bi-fundamental hypermultiplets.  The factors in the denominators denote the contributions of the $F$-terms associated with the middle $U(2)$ gauge node and the top left $U(1)$ gauge node.  Evaluating the integrals, we find the following refined and unrefined Hilbert series
\bea
H^{\text{Higgs}}_{\eref{quivD4s}, \text{left}} (t, y) &= \frac{1+t^6}{(1-t^4)^2} \times \PE \left[ \chi^{SU(2)}_{[2]}(y) t^2 -t^4\right]  \nn \\
&= \PE[\chi^{SU(2)}_{[2]}(y) t^2 +t^4 +t^6-t^{12} ]~, \\
H^{\text{Higgs}}_{\eref{quivD4s}, \text{left}} (t, y=1) &= \frac{1 - t^2 + t^4}{(1 - t^2)^4} \nn \\
& = 1 + 3 t^2 + 7 t^4 + 14 t^6 + 25 t^8 + 41 t^{10} +\ldots~.
\eea
We can therefore conclude that
\bea \label{C2D4C2Z2}
&\text{The Higgs branch of leftmost quiver in \eref{quivD4s} for $k=1$}  \nn \\
&=  (\BC^2/\widehat{D}_4) \times (\BC^2/\BZ_2) = (\BC^2/\widehat{D}_4) \times \CN_{SU(2)} ~,
\eea 
where $\BC^2/\widehat{D}_4$ is the orbifold singularity in question and $\BC^2/\BZ_2$ is the reduced moduli space of 1 $SU(2)$ instanton on $\BC^2$, which is the minimal nilpotent orbit $\CN_{SU(2)}$ of $SU(2)$.  It is interesting to compare this with the Coulomb branch of the same quiver, which is isomorphic to  $(\BC^2/\BZ_2) \times \CN_{SO(8)}$ (see \eref{C2Z2timesnilso8}). Notice that the group associated with the orbifold and that associated with the nilpotent orbit get exchanged. \\~\\
\noindent{\it The rightmost quiver of \eref{quivD4s}.} The Higgs branch of the rightmost diagram of \eref{quivD4s} has an interesting feature that worth discussing in detail.  Some properties of this space for $k=1$ were actually studied in section 4 of \cite{Lindstrom:1999pz}. In the following, we compute the Higgs branch Hilbert series of such a diagram for $k=1$.\footnote{This Hilbert series can alternatively be computed from \eref{fivefinger}, which is constructed by gluing 5 copies of $U(1)$ gauge theory with 2 flavours via the common $U(2)$ flavour symmetry, gauging this $U(2)$ symmetry and decoupling the overall $U(1)$ from this node.  The computation involves putting together 5 copies of $\BC^2/\BZ_2$, which is the Higgs branch of $U(1)$ gauge theory with 2 theory, and then modding out by the $U(2)/U(1)$ symmetry in the middle node:
\bea
H^{\text{Higgs}}_{\eref{quivD4s}, \text{right}} (t) &= \oint_{|z|=1} \frac{dz}{2 \pi i z} (1-z^2) \frac{\Big( \PE [\chi^{SU(2)}_{[2]}(z) t^2 -t^4] \Big)^5}{\PE[\chi^{SU(2)}_{[2]}(z) t^2]} = \frac{1-2 t^2+4 t^4-2 t^6+t^8}{\left(1-t^2\right)^4 \left(1+t^2\right)^2}~.
\eea}
\bea
H^{\text{Higgs}}_{\eref{quivD4s}, \text{right}} (t) &= \left( \prod_{i=1}^4 \oint_{|q_i|=1} \frac{dq_i}{2\pi i q_i} \right) \oint_{|z_1|=1}  \frac{dz_1}{2\pi i z_1} \oint_{|z_2|=1}  \frac{dz_2}{2\pi i z_2} (1-z_1 z_2^{-1})  \times \nn \\
& \qquad \PE \Big[ (z_1^{-1}+z_2^{-1})(q_1+\ldots q_4+1) t+(z_1+z_2)(q_1^{-1}+\ldots q_4^{-1}+1) t  \nn \\
& \qquad \qquad  - (z_1+z_2)(z_1^{-1}+z_2^{-1}) t^2 - 4t^2\Big] \nn\\
& = \frac{1-2 t^2+4 t^4-2 t^6+t^8}{\left(1-t^2\right)^4 \left(1+t^2\right)^2}  \nn\\
& = 1+5 t^4+4 t^6+15 t^8+16 t^{10}+\ldots~.
\eea
The fact that the coefficient of $t^2$ vanishes implies that the flavour symmetry is empty.  As a matter of fact, the Higgs branch of this quiver is isomorphic to the orbifold $\BC^4/\Gamma_{32}$, where $\Gamma_{32}$ refers to extraspecial group of order $32$ and type $-$ (also known as $\Gamma_5 a_2$) in \url{https://groupprops.subwiki.org/wiki/Central_product_of_D8_and_Q8}.  It is isomorphic to the central product of  the dihedral group $D_8$ of order $8$ and the quaternion group $Q_8$.\footnote{The group $\Gamma_5 a_2$ has a close cousin, namely the extraspecial group of order $32$ and type $+$ (also known as $\Gamma_5 a_1$). The latter is isomorphic to the central product of two dihedral groups of order 8, $D_8*D_8$, and to the central product of two quaternion groups, $Q_8 * Q_8$; see \url{https://groupprops.subwiki.org/wiki/Groups_of_order_32}. The Hilbert series for $\BC^4/\Gamma_5 a_1$ can be computed using the discrete Molien formula similarly to \eref{C4Gamma32}.  The result is
\be
\frac{1 - t^2 + 2 t^4 - t^6 + t^8}{(1 - t^2)^4 (1 + t^2)^2} = \frac{(1 + t^4) (1 - t^2 + t^4)}{(1 - t^2)^4 (1 + t^2)^2} = 1 + t^2 + 5 t^4 + 6 t^6 + 15 t^8 + 19 t^{10}+\ldots~.
\ee
Note that $\Gamma_5 a_2$ was misidentified as $\Gamma_5 a_1$ the previous versions of this article.  This is corrected in version 3 on the {\tt arXiv}. I thank Amihay Hanany and Travis Schedler for pointing this out.}.  In order to check this, we compute the Hilbert series of $\BC^4/\Gamma_{32}$ using the discrete Molien formula:\footnote{I thank Amihay Hanany and Martin Ro\v{c}ek for collaborating on this result during the Simons Workshop in Mathematics and Physics 2013.} 
\bea
H[\BC^4/\Gamma_{32}](t) = \frac{1}{32} \sum_{M \in \Gamma_{32}} \frac{1}{{\det( \bf 1}_{4 \times 4} - t M)}= \frac{1-2 t^2+4 t^4-2 t^6+t^8}{\left(1-t^2\right)^4 \left(1+t^2\right)^2}~, \label{C4Gamma32}
\eea
where the elements of $\Gamma_{32}$ are the Kronecker products $\mathbf{D}_i \otimes \mathbf{Q}_j$ (with $i,j=1, \ldots, 8$ labelling the elements of $\mathbf{D}$ and $\mathbf{Q}$ defined below) such that the repeated elements are counted once so that there are in total 32 elements.  The sets of matrices $\mathbf{D}$ and $\mathbf{Q}$ are defined as follows:
\bea
\mathbf{D} &= \{R_k, S_k \,:\, k=0,1,2,3 \}~, \\
\mathbf{Q} &= \{\mathbf{1}, \mathbf{I}, \mathbf{J}, \mathbf{K},- \mathbf{1}, -\mathbf{I}, -\mathbf{J},- \mathbf{K} \}~,
\eea
where
\bea
&R_k = \begin{pmatrix} \cos \, \frac{2 \pi k}{4} &\quad -\sin \, \frac{2 \pi k}{4} \\  \sin \, \frac{2 \pi k}{4} &\quad \cos \, \frac{2 \pi k}{4} \end{pmatrix}~, \qquad  S_k = \begin{pmatrix} \cos \, \frac{2 \pi k}{4} &\quad \sin \, \frac{2 \pi k}{4} \\  \sin \, \frac{2 \pi k}{4} &\quad -\cos \, \frac{2 \pi k}{4} \end{pmatrix}~, \\
& \mathbf{1} = \begin{pmatrix} 1 & 0 \\ 0 & 1\end{pmatrix}~, \quad \mathbf{I} = \begin{pmatrix} 0 & -1 \\ 1 & 0\end{pmatrix}~, \quad \mathbf{J} = \begin{pmatrix} 0 & i \\ i & 0\end{pmatrix}~, \quad \mathbf{K} = \begin{pmatrix} -i & 0 \\ 0 & i\end{pmatrix}~.
\eea
The plethystic logarithm of \eref{C4Gamma32} is
\bea
\PL \left[ H[\BC^4/\Gamma_{32}](t) \right] = 5 t^4 + 4 t^6 - 4 t^{10} - 10 t^{12} + \ldots~,
\eea
indicating that there are 5 generators at order $t^4$ and 4 generators at order $t^6$ subject to $4$ relations at order $t^{10}$ and $10$ relations at order $t^{12}$.

\subsubsection{Flavoured $B_3$ affine Dynkin diagrams}   
Below we present the generalised quivers diagrams whose Coulomb branches correspond to the moduli space of $k$ $SO(7)$ instantons on $\BC^2/\BZ_2$ with various monodromies at infinity specified by the residual symmetry of $SO(7)$. The symmetry of the Coulomb branch of each quiver can be read off from the Hilbert series computed using the prescription given in \cite{Cremonesi:2014xha}. For each quiver, $SO(7)$ symmetry is broken to $SO(7)$, $SO(5) \times SO(2)$, $SO(3) \times SO(4)$ and $O(1) \times SO(6)$, respectively from left to right and top to bottom.  The factor of $SU(2)$ that appears under each diagram corresponds to the isometry of $\BC^2/\BZ_2$.

\bea \label{quivB3s}
\begin{array}{ccc}
\begin{tikzpicture}[align=center,node distance=0.8cm]
  \node[draw, circle] (2) at (0,0) {{\footnotesize $2k$}};
  \node[draw, circle, below left=of 2] (0) {{\footnotesize $k$}};
  \node[draw, circle, above left=of 2] (1) {{\footnotesize $k$}};
  \node[draw, circle, right= of 2] (3) {{\footnotesize $k$}};
  \node[draw, rectangle, left= of 0] (flv) {{\footnotesize $2$}};
  \draw[-] (0) to (2);
  \draw[-] (1) to (2);
   \draw[double distance=3pt,arrows={-latex'}] (2) to (3);
\draw[-] (flv) to (0);
\node[draw=none] at (0,-2) {{\footnotesize $SU(2) \times SO(7)$}};
\end{tikzpicture} 
& \qquad  \qquad
\begin{tikzpicture}[align=center,node distance=0.8cm]
  \node[draw, circle] (2) at (0,0) {{\footnotesize $2k$}};
  \node[draw, circle, below left=of 2] (0) {{\footnotesize $k$}};
  \node[draw, circle, above left=of 2] (1) {{\footnotesize $k$}};
  \node[draw, circle, right= of 2] (3) {{\footnotesize $k$}};
  \node[draw, rectangle, left= of 0] (flv1) {{\footnotesize $1$}};
  \node[draw, rectangle, left= of 1] (flv2) {{\footnotesize $1$}};
  \draw[-] (0) to (2);
  \draw[-] (1) to (2);
   \draw[double distance=3pt,arrows={-latex'}] (2) to (3);
\draw[-] (flv1) to (0);
\draw[-] (flv2) to (1);
\node[draw=none] at (0,-2) {{\footnotesize $SU(2) \times SO(5) \times SO(2)$}};
\end{tikzpicture} 
\\
\begin{tikzpicture}[align=center,node distance=0.8cm]
  \node[draw, circle] (2) at (0,0) {{\footnotesize $2k$}};
  \node[draw, circle, below left=of 2] (0) {{\footnotesize $k$}};
  \node[draw, circle, above left=of 2] (1) {{\footnotesize $k$}};
  \node[draw, circle, right= of 2] (3) {{\footnotesize $k$}};
  \node[draw, rectangle] (flv) at (0,-1) {{\footnotesize $1$}};
  \draw[-] (0) to (2);
  \draw[-] (1) to (2);
   \draw[double distance=3pt,arrows={-latex'}] (2) to (3);
\draw[-] (flv) to (2);
\node[draw=none] at (0,-2) {{\footnotesize $SU(2) \times SO(3) \times SO(4)$}};
\end{tikzpicture} 
& \qquad \qquad
\begin{tikzpicture}[align=center,node distance=0.8cm]
  \node[draw, circle] (2) at (0,0) {{\footnotesize $2k$}};
  \node[draw, circle, below left=of 2] (0) {{\footnotesize $k$}};
  \node[draw, circle, above left=of 2] (1) {{\footnotesize $k$}};
  \node[draw, circle, right= of 2] (3) {{\footnotesize $k$}};
  \node[draw, rectangle, right= of 3] (flv) {{\footnotesize $1$}};
  \draw[-] (0) to (2);
  \draw[-] (1) to (2);
   \draw[double distance=3pt,arrows={-latex'}] (2) to (3);
\draw[-] (flv) to (3);
\node[draw=none] at (1,-2) {{\footnotesize $SU(2) \times O(1) \times SO(6)$}};
\end{tikzpicture} 
\end{array}
\eea

Let us now compute the Coulomb branch Hilbert series of the above quivers.  For the sake of brevity in writing the formulae below, let us define
\bea
&H[B_3](\Delta_{m}) \nn \\
&=\sum_{u_0 \in \BZ} \; \sum_{u_1 \in \BZ}\; \sum_{n_1 \geq n_2 > -\infty} \; \sum_{u_3 \in \BZ} t^{\Delta_m-2|n_1-n_2|+\sum_{i=1}^2 |u_0-n_i|+ |u_1-n_i|+|2n_i-u_3|} \times \nn \\
& \qquad z_0^{u_0} z_1^{u_1} z_2^{n_1+n_2} z_2^{u_3} \; (1-t^2)^{-3} P_{U(2)}(t; n_1, n_2) ~,
\eea
where $\Delta_m$ denotes the contribution from the fundamental matter attached to the $B_3$ affine quiver.  It may depend on $u$'s and $n$'s.

The Coulomb branch Hilbert series of the top left quiver in \eref{quivB3s} is
\bea
&H^{\text{Coul}}_{\eref{quivB3s}, \text{top left}} (t; z_0, \ldots, z_4) \nn \\
&= H[B_3](2|u_0|) \nn \\
&= \left( \sum_{p_1=0}^\infty \chi^{SU(2)}_{[2p_1]}(x) t^{2p_1} \right)\left( \sum_{p_2=0}^\infty \chi^{SO(7)}_{[0,p_2,0,0]}(y_1, \ldots, y_4) t^{2p_2} \right)~,
\eea
where
\bea
x^2= z_0 z_1 z_2^2 z_3^2~, \qquad y_1= z_1,  \qquad y_2= z_2, \qquad y_3 = z_3, \qquad y_4=z_4~.
\eea
This is the Hilbert series of $(\BC^2/\BZ_2) \times \CN_{SO(7)}$, where $\CN_{SO(7)}$ is  the minimal nilpotent orbit of $SO(7)$.

The Coulomb branch Hilbert series of the top right quiver in \eref{quivB3s} is
\bea
&H^{\text{Coul}}_{\eref{quivB3s}, \text{top right}} (t; z_0, \ldots, z_4) \nn \\
&= H[B_3](|u_0|+|u_1|) \nn \\
&= \PE \Big[ \Big(\chi^{SU(2)}_{[2]}(x)+\chi^{SO(5)}_{[0,2]}(\vec y) + 1\Big)t^2 +(\chi^{SU(2)}_{[1]} \chi^{SO(2)}_{[1]}(w) \chi^{SO(5)}_{[1,0]} (\vec y) ) t^3 \nn \\
& \qquad \qquad -(\chi^{SO(5)}_{[1,0]}(\vec y) + 2)t^4+ \ldots \Big]
\eea
where
\bea
x^2= z_0 z_1 z_2^2 z_3^2~, \qquad w_1^2= z_0 z_1^{-1}~, \qquad y_1 =z_2~, \qquad y_2 = z_3~, \qquad y_3 =z_4~.
\eea

The Coulomb branch Hilbert series of the bottom left quiver in \eref{quivB3s} is
\bea
&H^{\text{Coul}}_{\eref{quivB3s}, \text{bttm left}} (t; z_0, \ldots, z_4) \nn \\
&= H[B_3](|n_1|+|n_2|) \nn \\
&= \PE \Big[ \Big(\chi^{SU(2)}_{[2]}(x)+\chi^{SO(3)}_{[1]}(w)+\chi^{SO(4)}_{[2;0]+[0,2]}(\vec y) \Big)t^2 +(\chi^{SU(2)}_{[1]} \chi^{SO(2)}_{[1]}(w) \chi^{SO(5)}_{[1,0]} (\vec y) ) t^3 \nn \\
& \qquad \qquad -(\chi^{SO(5)}_{[1,0]}(\vec y) + 2)t^4+ \ldots \Big]
\eea
where
\bea
x^2= z_0 z_1 z_2^2 z_3^2~, \qquad w_1^2= z_0 z_1^{-1}~, \qquad y_1 =z_2~, \qquad y_2 = z_3~, \qquad y_3 =z_4~.
\eea

The Coulomb branch Hilbert series of the bottom right quiver in \eref{quivB3s} is
\bea
&H^{\text{Coul}}_{\eref{quivB3s}, \text{bttm right}} (t; z_0, \ldots, z_4) \nn \\
&= H[B_3](|u_3|) \nn \\
&=  \PE \Big[ \Big(\chi^{SU(2)}_{[2]}(x)+\chi^{SO(6)}_{[0,1,1]}(\vec y) \Big)t^2 +(\chi^{SU(2)}_{[1]} \chi^{SO(6)}_{[1,0,0]} (\vec y) ) t^3 \nn \\
& \qquad \qquad -(\chi^{SO(6)}_{[0,1,1]}(\vec y) + 2)t^4+ \ldots \Big]~,
\eea
where
\bea
x^2= z_0 z_1 z_2^2 z_3^2~, \qquad y_1 =z_1~, \qquad y_2 = z_2~, \qquad y_3 =z_0^{-1} z_1^{-1} z_2^{-1}~.
\eea

The above moduli spaces of instantons can be realised from the {\it Higgs branch} of the following quiver diagram (see section 4.1 of \cite{Dey:2013fea}):
\bea
\begin{tikzpicture}[baseline, align=center, node distance=0.8cm, minimum size=1.2cm]
  \node[draw, rectangle] (1) at (0,0) {{\footnotesize $SO(p)$}};
  \node[draw, circle, right=of 1] (2) {{\scriptsize $USp(2k)$}};
  \node[draw, circle, right=of 2] (3) {{\scriptsize $USp(2k)$}};
  \node[draw, rectangle, right= of 3] (4) {{\footnotesize $SO(7-p)$}};
  \draw[-] (1) to (2);
  \draw[-] (2) to (3);
  \draw[-] (3) to (4);
\end{tikzpicture} 
\eea
where $p=0,2,4,6$.  The symmetry of the Higgs branch of this theory is $SU(2) \times (SO(p) \times SO(7-p))$, where the $SU(2)$ factor is associated with the bifundamental hypermultiplet under $USp(2k) \times USp(2k)$; this is in agreement with those written in \eref{quivB3s}.  The Higgs branch Hilbert series of this quiver can be computed as in \cite{Dey:2013fea}.  This yields the same results as the Coulomb branch Hilbert series of \eref{quivB3s}, and hence provides a non-trivial check of the proposal in \cite{Cremonesi:2014xha}.

\subsubsection{Flavoured $G_2$ affine Dynkin diagrams}  
Below we present the generalised quivers diagrams whose Coulomb branches correspond to the moduli space correspond to $k$ $G_2$ instantons on $\BC^2/\BZ_2$ with various monodromies at infinity. For each quiver, the monodromy at infinity breaks $G_2$ symmetry to $G_2$ and to $SU(2) \times SU(2)$, respectively from left to right.  These can be seen from the Coulomb branch Hilbert series computed below.
\tikzset{
doublearrow/.style={draw, thin, double distance=4pt, arrows={-latex'}},
thirdline/.style={draw, thin, arrows={-latex'}}
}
\bea \label{1G2instonC2Z2}
\begin{tikzpicture}[baseline, align=center, node distance=0.8cm]
  \node[draw, rectangle] (1) at (0,0) {{\footnotesize $2$}};
  \node[draw, circle, right=of 1] (2) {{\scriptsize $k$}};
  \node[draw, circle, right=of 2] (3) {{\scriptsize $2k$}};
  \node[draw, circle, right= of 3] (4) {{\footnotesize $k$}};
  \draw[-] (1) to (2);
  \draw[-] (2) to (3); 
  \draw[-] (3) to (4);
  \path[doublearrow] (3) to (4);
  \path[thirdline] (3) to (4);
  \node[draw=none] at (2,-1) {{\footnotesize $SU(2) \times G_2$}};
\end{tikzpicture} 
\qquad \qquad \quad
\begin{tikzpicture}[baseline, align=center, node distance=0.8cm]
  \node[draw, circle, right=of 1] (2) at (0,0) {{\scriptsize $k$}};
  \node[draw, circle, right=of 2] (3) {{\scriptsize $2k$}};
  \node[draw, circle, right= of 3] (4) {{\footnotesize $k$}};
  \node[draw, rectangle, below=of 3] (1) {{\footnotesize $1$}};
  \draw[-] (1) to (3);
  \draw[-] (2) to (3); 
  \draw[-] (3) to (4);
  \path[doublearrow] (3) to (4);
  \path[thirdline] (3) to (4);
  \node[draw=none] at (2.5,-2.2) {{\footnotesize $SU(2) \times (SU(2) \times SU(2))$}};
\end{tikzpicture} 
\eea
For the sake of brevity in writing the formulae below, let us define
\bea
&H[G_2](\Delta_{m}) \nn \\
&=\sum_{u_0 \in \BZ}\; \sum_{n_1 \geq n_2 > -\infty}  \sum_{u_2 \in \BZ} t^{\Delta_m-2|n_1-n_2|+\sum_{i=1}^2 |u_0-n_i|+|3n_i-u_2|} \times \nn \\
& \qquad z_0^{u_0} z_1^{n_1+n_2} z_2^{u_2} \; (1-t^2)^{-2} P_{U(2)}(t; n_1, n_2) ~,
\eea
where $\Delta_m$ denotes the contribution from the fundamental matter attached to the $G_2$ affine quiver.  It may depend on $u$'s and $n$'s.

The Coulomb branch Hilbert series of the left quiver in \eref{1G2instonC2Z2} can be computed using the prescription given in \cite{Cremonesi:2014xha} as follows:
\bea
H^{\rm Coul}_{\eref{1G2instonC2Z2},\; \text{left}} (t; z_1, z_2, z_3) & = H[G_2](2|u_0|)\nn \\
&= \left( \sum_{p_1=0}^\infty \chi^{SU(2)}_{[2p_1]} (x) t^{2p_1} \right) \left( \sum_{p_2=0}^\infty \chi^{G_2}_{[0,p_2]} (y_1, y_2) t^{2p_2} \right)~,
\eea
where we donate by $[0,1]$ the adjoint representation of $G_2$, and
\bea
x^2 = z_0 z_1^2 z_2^3 , \qquad  y_1 = z_1, \qquad  y_2 =z_2~.
\eea
Indeed, this is the Hilbert series of $\BC^2/\BZ_2 \times \CN_{G_2}$, where $\CN_{G_2}$ denotes the minimal nilpotent orbit of $G_2$.

Similarly, the Coulomb branch Hilbert series of the right quiver in \eref{1G2instonC2Z2} is
\bea
H^{\rm Coul}_{\eref{1G2instonC2Z2},\; \text{right}} (t; z_1, z_2, z_3) &= H[G_2](|n_1|+|n_2|) \nn \\
&= \PE \Big[  \left( [2;0;0]+[0;2;0]+[0;0;2]\right)t^2 +  [1;1;3]t^3 \nn \\
& \qquad + ([0;2;6]-2) t^4 +\ldots \Big]~,
\eea
where
\bea
[a;b;c] = \chi^{SU(2)}_{[a]} (x) \chi^{SU(2)}_{[b]} (y_1)\chi^{SU(2)}_{[c]} (y_2)~,
\eea
and
\bea
x^2 = z_0 z_1^2 z_2^3, \qquad y_1^2 = z_1, \qquad y_2^2 = z_2~.
\eea


\section{Instantons on a smooth ALE space} \label{sec:smoothALE}
Hitherto we have discussed the moduli spaces of instantons on orbifold singularities and their $3d$ field theory realisations.  In this section, we consider the situation in which such singularities are resolved and the study moduli space of instantons on the smooth ALE space.  
The goal is to describe such a space using the moduli spaces of $3d$ $\CN=4$ gauge theories.

For instantons in a unitary gauge group on the ALE space of type $A_{n-1}$, the blow-up parameters of $\BC^2/\BZ_n$ are given by the Fayet-Iliopoulos (FI) terms \cite{Douglas:1996sw} of the KN quiver \eref{KN}.  Under mirror symmetry, these FI parameters are in correspondence with the mass parameters \cite{Intriligator:1996ex} in the mirror gauge theory described around \eref{CoulombKNquivconf}.

On the other hand, for instantons in an orthogonal gauge group, one may not be able to turn on such an FI term, for example, in quiver \eref{SOUSpUSpSO}.  Thus, the blow-up parameter may not be apparent from the weakly coupled Lagrangian description.  In particular, for $SO(8)$ instantons on the smooth ALE space $\widetilde{\BC^2/\BZ_{2n}}$ with monodromy at infinity such that $SO(8)$ is broken to $SO(4) \times SO(4)$, it was pointed out in \cite{Tachikawa:2014qaa} that the blow-up parameter of $\widetilde{\BC^2/\BZ_{2n}}$ can be made apparent in the field theoretic description by exploiting theories of class $\CS$ and their dualities.  

The purpose of this section is to reformulate such a description using $3d$ gauge theories and present a quiver whose Coulomb branch describes the moduli space of $SO(2N)$ instantons on the smooth ALE space $\widetilde{\BC^2/\BZ_{2n}}$.  For simplicity, we restrict our presentation to the case in which the monodromy at infinity breaks $SO(2N)$ to $SO(2N-4) \times SO(4)$.  For $N=4$, we compare our results with those in \cite{Tachikawa:2014qaa}, where a different approach was adopted.  In fact, it was proven in \cite{kronheimer1990yang} that the Higgs branch of such quiver describes the moduli space of $PU(2n)$ instantons on $\widetilde{\BC^2/\widehat{D}_{N}}$ with monodromy at infinity such that $PU(2n)$ is unbroken.

\subsection{$SO(8)$ instantons on $\widetilde{\BC^2/\BZ_2}$} \label{sec:SO8smoothC2Z2}
Let us start from the case of $k$ $SO(8)$ instantons on $\widetilde{\BC^2/\BZ_2}$ with the monodromy at infinity such that $SO(8)$ is broken to $SO(4) \times SO(4)$.  The moduli space of this configuration of instantons is identified with the {\it Higgs branch} of the following quiver (see section 4.2 of \cite{Intriligator:1997kq} and  section 2 of \cite{Tachikawa:2014qaa}):
\bea \label{SO8smoothC2Z2}
\begin{tikzpicture}[baseline, align=center, node distance=1cm,minimum size=1cm]
  \node[draw, rectangle] (1) at (0,0) {{\footnotesize $SO(4)$}};
  \node[draw, ellipse, right=of 1] (2) {{\scriptsize $USp(2k)$}};
  \node[draw, ellipse, right=of 2] (3) {{\scriptsize $USp(2k+2)$}};
  \node[draw, rectangle, right= of 3] (4) {{\footnotesize $SO(4)$}};
  \draw[-] (1) to (2);
  \draw[-] (2) to (3);
  \draw[-] (3) to (4);
\end{tikzpicture} 
\eea
Note that this $3d$ $\CN=4$ quiver  is a {\it bad} theory in the sense of \cite{Gaiotto:2008ak} for all $k\geq 1$, because the $USp(2k+2)$ gauge group has $2k+2$ flavours of fundamental hypermultiplets charged under it\footnote{A linear quiver is a  bad theory if it contains a $USp(2N_c)$ gauge group with $N_f <2N_c+1$ flavours.}.   Motivated by \eref{quivD4s} and by the brane configuration in \cite{Hanany:1999sj}, we propose that a mirror theory of quiver \eref{SO8smoothC2Z2} is
\bea 
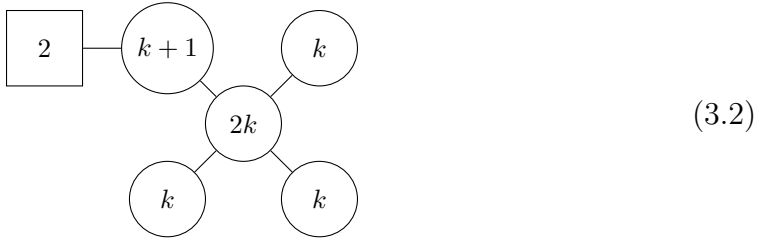
\label{mirrSO8smoothC2Z2}
\begin{tikzpicture}[baseline, align=center,node distance=0.5cm,minimum size=1cm]
  \node[draw, circle] (4) at (0,0) {{\footnotesize $2k$}};
  \node[draw, circle] (0) at (-1,1) {{\footnotesize $k+1$}};
  \node[draw, circle] (1) at (-1,-1) {{\footnotesize $k$}};
  \node[draw, circle] (2) at (1,-1) {{\footnotesize $k$}};
  \node[draw, circle] (3) at (1,1) {{\footnotesize $k$}};
  \node[draw, rectangle, left= of 0] (flv) {{\footnotesize $2$}};
\foreach \s in {0,...,3}
{
  \draw[-] (4) to (\s);
}
\draw[-] (flv) to (0);
\end{tikzpicture} 
\eea
Thus, the Coulomb branch of \eref{mirrSO8smoothC2Z2} describes the moduli space of $SO(8)$ instantons on $\widetilde{\BC^2/\BZ_2}$ with monodromy at infinity such that $SO(8)$ is broken to $SO(4) \times SO(4)$.

The quaternionic dimension of the Coulomb branch of \eref{mirrSO8smoothC2Z2} is 
\bea
(k+1)+k+k+k+2k= 6k+1~, 
\eea
and the quaternionic dimension of the Higgs branch of \eref{mirrSO8smoothC2Z2} is 
\bea
3(2k)(k)+2k(k+1)+2(k+1)-[3k^2+(2k)^2+(k+1)^2]=2k+1~.
\eea  
The former is indeed the dimension of the moduli space of $k$ $SO(8)$ instantons on $\widetilde{\BC^2/\BZ_2}$ \cite{Intriligator:1997kq, Tachikawa:2014qaa}, and the latter is equal to with the total rank of the gauge groups in quiver \eref{SO8smoothC2Z2}.  

In quiver \eref{mirrSO8smoothC2Z2}, there are two flavours of fundamental hypermultiplets charged under gauge group $U(k+1)$.  Hence the theory admits one mass parameter that cannot be eliminated by shifting any of the dynamical fields (this is actually the difference between the masses of the two flavours).  This mass parameter is indeed identified with the blow-up parameter of $\widetilde{\BC^2/\BZ_2}$.  Note that, under mirror symmetry, this mass parameter corresponds to a ``Fayet-Iliopoulos parameter'' that is not visible in the Lagrangian of \eref{SO8smoothC2Z2}; this is an example of the ``hidden'' FI parameter discussed in \cite{Kapustin:1998fa}.

It is also instructive to compare \eref{fivefinger} with \eref{mirrSO8smoothC2Z2}.  Recall that the Coulomb branch of former describes the moduli space of $k$ $SO(8)$ instantons on $\BC^2/\BZ_2$.  Observe that theory \eref{fivefinger} does not admit a (non-trivial) mass parameter, whereas theory \eref{mirrSO8smoothC2Z2} admits a mass parameter that corresponds to the blow-up parameter of $\widetilde{\BC^2/\BZ_2}$.

Let us now compute the Coulomb branch Hilbert series of \eref{mirrSO8smoothC2Z2}.  The computation can be performed using the method described in \cite{Cremonesi:2013lqa} and Appendix \ref{app:Coulomb}. For brevity, we report only the unrefined Hilbert series for $k=1$ and $k=2$ here:
\bea
H^{\text{Coul}}_{\eref{mirrSO8smoothC2Z2}, \;  k=1}(t) \label{CoulHSk1}
&=\frac{1}{(1 - t)^{14} (1 + t)^6 (1 + t^2)^3 (1 + t + t^2)^7} \times \Big( 1 - t + 12 t^2 + 13 t^3 + 56 t^4 \nn \\
& \quad + 123 t^5 + 248 t^6 + 453 t^7 + 
 777 t^8 + 1090 t^9 + 1580 t^{10} + 1890 t^{11} + 2182 t^{12} \nn \\
& \quad + 2280 t^{13} + 
 2182 t^{14} + \text{palindrome} + t^{26}\Big) \nn \\
&= 1+15 t^2+32 t^3+126 t^4+384 t^5+1025 t^6+\ldots~, \\
H^{\text{Coul}}_{\eref{mirrSO8smoothC2Z2}, \;  k=2}(t)&= 1 + 15 t^2 + 32 t^3 + 161 t^4 + 544 t^5 + 1820 t^6 +\ldots~.
\eea
Note that the coefficient $15$ of $t^2$ is the sum of the dimensions of the adjoint representations of each factor in the symmetry of the Coulomb branch $SU(2) \times SO(4) \times SO(4)$, where $SU(2)$ corresponds to the isometry of $\widetilde{\BC^2/\BZ_2}$.

For $k=1$, theory \eref{mirrSO8smoothC2Z2} has another mirror theory which is a good theory in the sense of \cite{Gaiotto:2008ak}:
\bea 
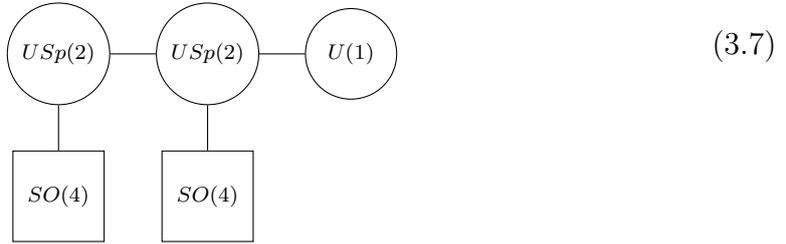
\label{anothermirrork1}
\begin{tikzpicture}[baseline, align=center, node distance=0.6cm,minimum size=1.2cm]
  \node[draw, circle] (1) {{\scriptsize $USp(2)$}};
  \node[draw, circle, right=of 1] (2) {{\scriptsize $USp(2)$}};
  \node[draw, circle, right=of 2] (3) {{\scriptsize $U(1)$}};
  \node[draw, rectangle, below= of 1] (4) {{\scriptsize $SO(4)$}};
  \node[draw, rectangle, below= of 2] (5) {{\scriptsize $SO(4)$}};
  \draw[-] (1) to (2) to (3);
  \draw[-] (1) to (4);
  \draw[-] (2) to (5);
\end{tikzpicture} 
\eea
This theory was actually pointed out in section 2.3 of \cite{Tachikawa:2014qaa} and the blow-up parameter of $\widetilde{\BC^2/\BZ_2}$ is indeed apparent in the Lagrangian description as the FI parameter of the $U(1)$ gauge group.  It can be checked that the Higgs branch Hilbert series of \eref{anothermirrork1} is equal to \eref{CoulHSk1}.

\subsection{$SO(2N)$ instantons on $\widetilde{\BC^2/\BZ_{2n}}$} \label{sec:SO2NC2Z2n}
It is straightforward to generalise the previous result to a general $SO(2N)$ gauge group and to a general smooth ALE space of type $A_{2n-1}$.  

We propose a quiver theory whose Coulomb branch describes the moduli space of $k$ $SO(2N)$ instantons on $\widetilde{\BC^2/\BZ_{2n}}$ with monodromy at infinity such that $SO(2N)$ is broken to $SO(2N-4) \times SO(4)$ to be as follows:
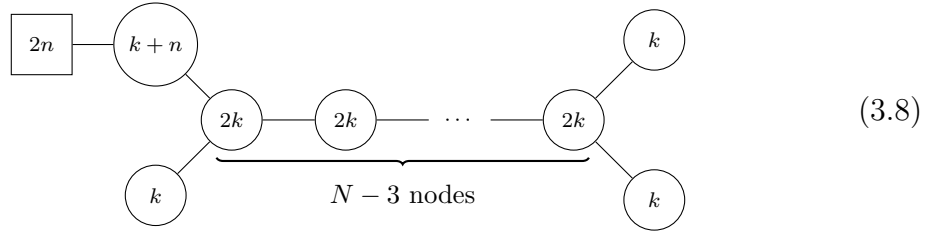
\bea \label{mirrSO2NinstC2Z2n}
\begin{tikzpicture}[baseline, align=center,node distance=1.5cm,minimum size=0.8cm]
  \node[draw, circle] (0) at (-1,1) {{\scriptsize $k+n$}};
  \node[draw, circle] (1) at (-1,-1) {{\scriptsize $k$}};
   \node[draw, circle] (4) at (0,0) {{\scriptsize $2k$}};
   \node[draw, circle, right of=4] (5)  {{\scriptsize $2k$}};
   \node[draw=none, right of=5] (6)  {{\scriptsize $\cdots$}};
   \node[draw, circle, right of=6] (7)  {{\scriptsize $2k$}};
    \node[draw, circle, above right of=7] (8) {{\scriptsize $k$}};
  \node[draw, circle, below right of =7] (9) {{\scriptsize $k$}};
  \node[draw, rectangle] (flv) at (-2.5,1) {{\scriptsize $2n$}};
\foreach \s in {0,...,1}
{
  \draw[-] (4) to (\s);
}
\foreach \s in {8,...,9}
{
  \draw[-] (7) to (\s);
}
\draw[-] (flv) to (0);
\draw[-] (4) to (5) to (6) to (7);
\draw [
    thick,
    decoration={
        brace,
        mirror,
        raise=0.5cm
    },
    decorate
] (-0.2,0) -- (4.7,0);
\node[draw=none] at (2.25,-1) {{\footnotesize $N-3$ nodes}};
\end{tikzpicture} 
\eea
We claim that this theory is a $3d$ mirror of the following quiver
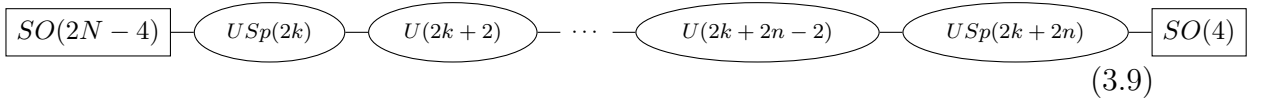
\bea \label{SO2NsmoothC2Z2n}
\begin{tikzpicture}[baseline, align=center, node distance=0.3cm,minimum size=0.5cm]
  \node[draw, rectangle] (1) at (0,0) {{\footnotesize $SO(2N-4)$}};
  \node[draw, ellipse, right=of 1] (2) {{\scriptsize $USp(2k)$}};
  \node[draw, ellipse, right=of 2] (3) {{\scriptsize $U(2k+2)$}};
   \node[draw=none, right=of 3] (4) {{\scriptsize $\cdots$}};
   \node[draw, ellipse, right=of 4] (5) {{\scriptsize $U(2k+2n-2)$}};
   \node[draw, ellipse, right=of 5] (6) {{\scriptsize $USp(2k+2n)$}};
  \node[draw, rectangle, right= of 6] (7) {{\footnotesize $SO(4)$}};
  \draw[-] (1) to (2) to (3) to (4) to (5) to (6) to (7);
\end{tikzpicture} 
\eea
Note that, for $N=4$, quiver \eref{SO2NsmoothC2Z2n} was discussed in Figure 9 of \cite{Tachikawa:2014qaa}.

Notice that the quaternionic dimension of the Coulomb branch of \eref{mirrSO2NinstC2Z2n} is 
\bea
{\rm dim}_{\BH} ~ \text{Coulomb of \eref{mirrSO2NinstC2Z2n}} =  (2N-2)k+n = k h^\vee_{SO(2N)} +n ~, 
\eea
as expected. The quaternionic dimension of the Higgs branch of \eref{mirrSO2NinstC2Z2n} is 
\bea \label{dimHiggsmirr}
{\rm dim}_{\BH} ~ \text{Higgs of \eref{mirrSO2NinstC2Z2n}} = 2nk+n^2 = k h^\vee_{SU(2n)} +n^2 ~,
\eea
in agreement with the dimension of the Coulomb branch of \eref{SO2NsmoothC2Z2n}.  Note that the latter is independent of $N$.

Theory \eref{mirrSO2NinstC2Z2n} admits $2n-1$ non-trivial mass parameters for the $2n$ flavours of the fundamental hypermultiplets under gauge group $U(k+n)$. These mass parameters are identified with the blow-up parameters of $\widetilde{\BC^2/\BZ_{2n}}$.

In order to check this proposal, let us compute the Coulomb branch Hilbert series of \eref{mirrSO2NinstC2Z2n} for some values of $k$, $N$ and $n$ using the method described in \cite{Cremonesi:2013lqa} and  Appendix \ref{app:Coulomb}.   The unrefined Hilbert series for $k=1$, $N=5$ and $n=1$ is given by
\bea \label{coulk1N5n1}
H^{\text{Coul}}_{\eref{mirrSO2NinstC2Z2n}, k=1, N=5, n=1}(t)=1 + 24 t^2 + 48 t^3 + 292 t^4 + 944 t^5 + 3279 t^6 +\ldots~.
\eea
The coefficient 24 of $t^2$ is the sum $3+15+6$ of the dimensions of the adjoint representations of each factor in the symmetry of the Coulomb branch $SU(2) \times SO(6) \times SO(4)$, where $SU(2)$ corresponds to the isometry of $\widetilde{\BC^2/\BZ_2}$.   We also check that the Coulomb branch Hilbert series \eref{coulk1N5n1} agrees with the Higgs branch Hilbert series of the following quiver
\bea
\begin{tikzpicture}[baseline, align=center, node distance=1cm,minimum size=1cm]
  \node[draw, rectangle] (1) at (0,0) {{\footnotesize $SO(6)$}};
  \node[draw, ellipse, right=of 1] (2) {{\scriptsize $USp(2)$}};
  \node[draw, ellipse, right=of 2] (3) {{\scriptsize $USp(4)$}};
  \node[draw, rectangle, right= of 3] (4) {{\footnotesize $SO(4)$}};
  \draw[-] (1) to (2);
  \draw[-] (2) to (3);
  \draw[-] (3) to (4);
\end{tikzpicture} 
\eea
Moreover, for $k=1$, $N=5$ and $n=2$, we find that the coefficient of $t^2$ in the unrefined Coulomb branch Hilbert series of \eref{mirrSO2NinstC2Z2n} is equal to $22$.  This is equal to the sum $1+15+6$ of the dimensions of the adjoint representations of each factor in the symmetry of the Coulomb branch $U(1) \times SO(6) \times SO(4)$, where $U(1)$ corresponds to the isometry of $\widetilde{\BC^2/\BZ_4}$.

\subsection{$PU(2n)$ instantons on $\widetilde{\BC^2/\widehat{D}_N}$} \label{sec:SU2nonsmoothC2DN}
It was proven in \cite{kronheimer1990yang} that the Higgs branch of \eref{mirrSO2NinstC2Z2n} describes the moduli space of $k$ $PU(2n)$ instantons on $\widetilde{\BC^2/\widehat{D}_N}$ with monodromy at infinity such that $PU(2n)$ remains unbroken.  As a result of mirror symmetry, the Coulomb branch of \eref{SO2NsmoothC2Z2n} also describes the same moduli space.  Indeed, the $N$ non-trivial FI parameters of the former and the $N$ non-trivial mass parameters in the latter can be identified with the blow-up parameters of $\widetilde{\BC^2/\widehat{D}_N}$.  




Let us examine the Higgs branch Hilbert series of \eref{mirrSO2NinstC2Z2n}.  For $k=1$, $n=1$ and $N=4$, namely one $PU(2)$ instanton on $\widetilde{\BC^2/\widehat{D}_4}$, the unrefined Hilbert series is
\bea \label{HiggsmirrSO2NinstC2Z2n}
H^{\text{Higgs}}_{\eref{mirrSO2NinstC2Z2n}}(t) &= \frac{1 + 2 t^4 + 4 t^6 + 4 t^8 + 4 t^{10} + 4 t^{12} + 2 t^{14} + t^{18}}{\left(1-t^2\right)^6 \left(1+t^2+t^4\right)^3} \nn \\
&= 1 + 3 t^2 + 8 t^4 + 23 t^6 + 52 t^8 + 105 t^{10} +204 t^{12} + 363 t^{14}+ \ldots \nn \\
&= \PE[ 3 t^2 + 2 t^4 + 7 t^6 + t^8 - 4 t^{10} - 12 t^{12} - 6 t^{14}+\ldots]~.
\eea
The coefficient 3 of $t^2$ is indeed the dimension of $PU(2) \cong U(2)/U(1)$.  It is worth contrasting this result with the the moduli space of one $PU(2)$ instanton on singular orbifold $\BC^2/\widehat{D}_4$ with monodromy at infinity such that $PU(2)$ is unbroken.  The latter is the product of two orbifolds $(\BC^2/\widehat{D}_4) \times (\BC^2/\BZ_2)$ according to \eref{C2D4C2Z2}, whereas the space corresponding to \eref{HiggsmirrSO2NinstC2Z2n} is not.


\section{Conclusions and open problems} \label{sec:conclude}
In this paper, we study $3d$ $\CN=4$ field theories whose Higgs branch and/or Coulomb branch describe the moduli space of instantons on a singular orbifold or on a smooth ALE space.  

In the first part of the paper, we show that for instantons in a simple gauge group $G$ on $\BC^2/\BZ_n$, the Hilbert series of such an instanton moduli space can be computed from the Coulomb branch of the quiver given by the affine Dynkin diagram of $G$ with flavour nodes of unitary groups attached to various nodes of the Dynkin diagram.  The techniques presented in \cite{Cremonesi:2013lqa, Cremonesi:2014xha} can thus be applied to compute the Coulomb branch Hilbert series of such quivers.  For $G$ a simply laced group of type $A$, $D$ or $E$, the Higgs branch of such a quiver describes the moduli space of $PU(n)$ instantons on $\BC^2/\widehat{G}$, where $\widehat{G}$ is the discrete group that is in McKay correspondence to $G$.  

In the second part of the paper, the moduli space of instantons on a smooth ALE space is discussed.  We present a quiver whose Coulomb branch describes the moduli space of $SO(2N)$ instantons on a smooth ALE space $\widetilde{\BC^2/\BZ_{2n}}$ with monodromy at infinity such that $SO(2N)$ is broken to $SO(2N-4) \times SO(4)$.  Various special cases of our results are checked against those presented in \cite{Tachikawa:2014qaa} and yield an agreement. We leave the study of other monodromies at infinity for future work.

Our work leaves a number of open questions.  In the case of singular orbifolds, one may naturally ask for a field theory that describes the moduli space of instantons in a non-unitary gauge group on orbifold $\BC^2/\Gamma$, where $\Gamma = \widehat{D}_n~\text{or}~\widehat{E}_{6,7,8}$.  In fact, a quiver description for the moduli space of $SO(2N)$ instantons on $\BC^2/\widehat{D}_n$ was proposed in Figure 17 in section 4.4 of \cite{Hanany:1999sj}.  However, upon considering the special case in which $SO(2N)$ remains unbroken by the monodromy at infinity (\ie~ when there is exactly one flavour node attached to the affine node of the Dynkin diagram), the quiver contains a $USp(2N_c)$ gauge group with $2N_c$ flavours.  This renders the quiver a bad theory in the sense of \cite{Gaiotto:2008ak} and all known methods for computing the Coulomb branch Hilbert series yield infinity.  Hence we cannot use the Coulomb branch Hilbert series to check that the Coulomb branch of the quiver in this case has a desired property, namely being either $\BC^2/\widehat{D}_{n} \times \CN_{SO(2N)}$ or $\BC^2/\widehat{D}_{N} \times \CN_{SO(2n)}$; cf. \eref{1pureSUN}, \eref{Higgs1pureSUN}, \eref{pure1G} and \eref{Higgspure1G}.   On the other hand, by computing the Higgs branch Hilbert series of such a quiver, we find that the Higgs branch does not have such a property either.  Given such a situation, we do not go into detail of this quiver and leave the study of such a moduli space of instantons for future work.

In the case of smooth ALE spaces, our knowledge is far from complete.  So far in this paper we have only provided a few examples that can be checked explicitly against the known results.  It would be nice to study such instanton moduli spaces in a systematic fashion.

\acknowledgments
I thank Stefano Cremonesi, Anindya Dey, Giulia Ferlito, Amihay Hanany, Peter Koroteev, Diego Rodriguez-Gomez and Alberto Zaffaroni for a close collaboration and invaluable insights over the years.  My sincere gratitude is extended to Stefano Cremonesi for encouraging me to write this paper and to Alberto Zaffaroni for reading the manuscript and for his insightful comments.  I am particularly indebted to Amihay Hanany and Martin Ro\v{c}ek for collaborating on the result around \eref{C4Gamma32} during the Simons Workshop in Mathematics and Physics 2013.  I acknowledges the following institutes, workshops and researchers for their hospitality and partial support: LPTHE at the University Pierre et Marie Curie (Jussieu), Nick Halmagyi; University of Rome Tor Vergata, Francesco Fucito, Francisco Morales and Massimo Bianchi; and the Simons Summer Workshop 2015.  I am also very grateful to Claudius Klare for his kind hospitality during my academic visit in Paris.  Special thanks go to Hiraku Nakajima for a useful discussion that contributes to version 2 of this paper.  I am indebted to Amihay Hanany and Travis Schedler for comments that have led to an improvement in version 3 of this paper.

\appendix
\section{The Coulomb branch Hilbert series} \label{app:Coulomb}
In this appendix, we review the computation of the Hilbert series for the Coulomb branch of $3d$ $\mathcal{N}=4$ quiver gauge theories whose gauge group is a product of unitary groups. As is discussed in the main text (and in \cite{Cremonesi:2014xha}), for suitable generalised quivers, possibly with non-simple laces, this method computes the Hilbert series of the moduli space of instantons.

A weak coupling description of a $3d$ $\mathcal{N}=4$ theory is specified by vector multiplets in the adjoint representation and matter fields (hypermultiplets or half-hypermultiplets) in some representation of the gauge group. At a generic point on the Coulomb branch the scalars in the vector multiplet acquire non-zero vacuum expectation values, breaking the gauge group $\cG$ of rank $r$ to $U(1)^r$.  As a result, matter fields and W-bosons gain masses and are integrated out, while the $r$ massless gauge fields, the photons, can be dualised to scalars.  Hence the low energy effective theory of the Coulomb branch consists of $r$ abelian vector multiplets which, by virtue of the gauge field dualisation to a scalar, can be themselves dualised to twisted hypermultiplets.

The previous description breaks down at the origin of the Coulomb branch, which corresponds to a strongly coupled superconformal fixed point in the infrared (IR).  Moreover, when the residual gauge group is non-abelian, the dualisation of a non-abelian vector multiplet is not understood.  A suitable description of the Coulomb branch at the fixed point involves disorder operators that cannot be described in terms of a polynomial in the microscopic degrees of freedom \cite{Borokhov:2002ib}.  These operators are known as {\it monopole operators} and are defined by specifying a Dirac monopole singularity at an insertion point in the Euclidean path integral \cite{'tHooft:1977hy}. Monopole operators are classified by embedding $U(1)$ into the gauge group $\cG$, and are labeled by magnetic charges which, by a generalised Dirac quantization \cite{Englert:1976ng}, take value in the weight lattice $\Gamma_{\cG^\vee}$ of the GNO or Langlands dual group $\cG^\vee$ \cite{Goddard:1976qe,Kapustin:2005py}. 
The monopole flux breaks the gauge group $\cG$ to a residual gauge group $H_{\vec m}$ by the adjoint Higgs mechanism. 
Restricting to gauge invariant monopole operators is achieved by modding out by the Weyl group $\cW_\cG$ of $\cG$, thus restricting $\vec m \in \Gamma_{\cG^\vee}/\cW_\cG$. 

In a $3d$ $\cN=2$ theory, half-BPS monopole operators are contained in the chiral multiplets. There is a \emph{unique} BPS monopole operator $V_{\vec  m}$ for each choice of magnetic charge $\vec m$ \cite{Borokhov:2002cg}. If the theory has $\cN=4$ supersymmetry, the $\cN=4$ vector multiplet decomposes into an $\cN=2$ vector multiplet $V$ and a chiral multiplet $\Phi$ in the adjoint representation. To describe the Coulomb branch, $V$ is replaced by monopole operators $V_{\vec m}$, which now can be dressed by the classical complex scalar $\phi$ inside $\Phi$. This dressing preserves the same supersymmetry of a chiral multiplet \cite{Borokhov:2003yu} if and only if $\phi$ is restricted to a constant element $\phi_{\vec m}$ of the Lie algebra of the residual gauge group $H_{\vec m}$ \cite{Cremonesi:2013lqa}. The monopole operators that parametrise the Coulomb branch of an $\cN=4$ field theory are therefore polynomials of $V_{\vec m}$ and $\phi_{\vec m}$, which are made gauge invariant by averaging over the action of the Weyl group \cite{Cremonesi:2013lqa}.  

Let us focus on a gauge group $\cG$ that is a product of $U(N_i)$ unitary groups, which are self-dual. For $U(N)$ monopole operators $V_{\vec m}$, with magnetic charge  $\vec m=\diag(m_1,...,m_N)$, the weight lattice of the dual group is given by $\Gamma_{U(N)}=\bZ^N = \left\{ m_i \in \BZ, i=1,..,N \right\}$. Modding out by the Weyl group $S_N$ restricts the lattice to the Weyl chamber $\Gamma_{U(N)}/S_N= \left\{\vec m \in \bZ^N| m_1 \geq m_2 \geq ... \geq m_N\right\}$.
For $U(N)$ gauge groups, which are not simply connected, its centre $U(1)$ corresponds to a topological $U(1)_J$ symmetry group. Classically, monopole operators are only charged under this symmetry. To each such $U(N_i)$ gauge group, we associate a fugacity $z_i$ for the topological $U(1)_{J_i}$ symmetry with conserved current $\ast \Tr F_i$, where $F_i$ is the field strength of the $i$-th gauge group. 

Other charges are acquired quantum-mechanically. In particular, monopole operators become charged under the 
Cartan $U(1)_C$ of the $SU(2)_C$ $R$-symmetry acting on the Coulomb branch. For a Lagrangian $\cN=4$ gauge theory with gauge group $\cG$, this charge is given by the formula 
\bea\label{dimension}
\Delta(\vec m)=-\sum_{{\vec \alpha} \in \Delta_{+}}{\left|\vec \alpha(\vec m) \right|} + \frac{1}{2} \sum_{i=1}^{n}{\sum_{{\vec \rho}_i \in \cR_{i}}{\left|{\vec \rho}_{i}(\vec m) \right|}} ~,
\eea
where the first contribution, arising from vector multiplets, is a sum over the positive roots of $\cG$, while the second contribution is a sum over the weights of representations of the gauge group $\cG$ of the hypermultiplets. The fugacity for this $R$-symmetry is called $t^2$ in the following. The dimension formula \eqref{dimension} was conjectured in \cite{Gaiotto:2008ak} based on a weak coupling computation in \cite{Borokhov:2002cg}, and later proven in \cite{Benna:2009xd,Bashkirov:2010kz}.
For the theories that we study in this paper, which are good or ugly in the sense of \cite{Gaiotto:2008ak}, \eqref{dimension} is believed to equal the scaling dimension in the IR CFT. 

The main focus of this paper is on field theories described by Dynkin diagrams, possibly non-simply laced and possibly with flavours. We propose the following prescription for computing the $R$-charge of a monopole operator, generalising the Lagrangian formula \eqref{dimension}. Each diagram is constructed from two basic building blocks: a node and a line.  They contribute to \eqref{dimension} as follows:
\bi
\item A $U(N)$ node, with magnetic charge $\vec m$, contributes to the Coulomb branch Hilbert series as follows:
\bea \label{dimension_vector}
\begin{tikzpicture}[baseline]
  \node[draw, circle] {{\tiny $U(N)$}};
\end{tikzpicture}
\qquad  \qquad 
\Delta_{\text{node}}(\vec m)= - \sum_{1\leq i < j \leq N}  
\left|m_i-m_j \right|~.
\eea
\item A line connecting the nodes $U(N_1)$ and $U(N_2)$ can be either a single line ($-$), a double line ($\Rightarrow$) or a triple line ($\Rrightarrow$), which we take to be oriented from node $1$ to node $2$. Let us assign magnetic charges $\vec m^{(1)}$ and $\vec m^{(2)}$ to $U(N_1)$ and $U(N_2)$ respectively. We propose that the contribution from a line is:
\bea \label{dimension_hyper_guess}
\begin{tikzpicture}[baseline, align=center, node distance=0.7cm]
\node[draw, circle] (nd1) at (0,0) {{\tiny $U(N_1)$}};
  \node[draw, circle, right=of nd1] (nd2) {{\tiny $U(N_2)$}};
    \draw[dashed]   (nd1)--(nd2) ;
	\end{tikzpicture}
\qquad 
\Delta_{\text{line}}(\vec m^{(1)},\vec m^{(2)})=\frac{1}{2}\sum_{j=1}^{N_1} \sum_{k=1}^{N_2}{\left|\lambda m^{(1)}_j-m^{(2)}_k \right|}
\eea
where $\lambda=1$ for a single bond, $\lambda=2$ for a double bond and $\lambda=3$ for a triple bond. If $\lambda>1$, \eqref{dimension_hyper_guess} does not arise from matter fields transforming in a genuine representation of $U(N_1)\times U(N_2)$.   

If one of the nodes, say $U(N_1)$, is a flavour node (\ie~ donated by a square), then the corresponding $\vec m^{(1)}$ should be treated a background magnetic flux (see \eg~ \cite{Cremonesi:2014kwa}).  In this paper, we turn off such background fluxes and hence the contribution to the dimension formula of a single line connecting a square node and a circular node is given by
\bea \label{dimension_hyper}
\begin{tikzpicture}[baseline, align=center, node distance=0.7cm, minimum size=1cm]
\node[draw, rectangle] (nd1) at (0,0) {{\tiny $U(N_1)$}};
  \node[draw, circle, right=of nd1] (nd2) {{\tiny $U(N_2)$}};
    \draw[-]   (nd1)--(nd2) ;
\end{tikzpicture}
\qquad 
\Delta_{\text{line}}(\vec m^{(1)}= \vec 0,\vec m^{(2)})=\frac{1}{2}N_1  \sum_{k=1}^{N_2}{\left|m^{(2)}_k \right|}~.
\eea
\ei
The dimension formula is then given by the sum of the two contributions, \eqref{dimension_vector} for each node and \eqref{dimension_hyper_guess} or \eref{dimension_hyper} for each line.  



Let us enumerate gauge invariant chiral operators on the Coulomb branch of non-simply laced quivers according to their quantum number $J_i$ and $\Delta$. The generating function, also known as the {\it Coulomb branch Hilbert series}, of such operators is given by  \cite{Cremonesi:2013lqa}
\be\label{ref_HS}
H(t,\vec{z})=\sum_{\vec{m}\in \Gamma_{\cG^\vee}/\cW_{\cG}} \vec{z}^{\vec{J}(\vec{m})}~ t^{2\Delta(\vec{m})} P_{\cG}(t;\vec{m})~,
\ee
where $\vec{z}^{\vec{J}(\vec{m})}=\prod_i z_i^{J_i(m)}$.  Each component of the formula can be explained as follows:
\bi
\item The sum is over GNO magnetic sectors \cite{Goddard:1976qe}, restricted to a Weyl chamber to impose invariance under the gauge group $\cG$. There is precisely one bare monopole operator per magnetic charge sector \cite{Borokhov:2002cg}.  
\item The factors $\vec{z}^{\vec{J}(\vec{m})}~ t^{2\Delta(\vec{m})}$ account for the topological charges and conformal dimension of bare monopole operators of magnetic charge $\vec{m}$. 
\item The factor $P_{\cG}(t;\vec{m})$ reflects the dressing of a bare monopole operator $V_{\vec{m}}$ by polynomials of the classical adjoint scalar $\phi_{\vec m}$ which are gauge invariant under the residual gauge group $H_{\vec{m}}$ left unbroken by the monopole flux. The contribution of this dressing factor to the Hilbert series is given by the generating function of independent Casimir invariants under the residual gauge group $H_{\vec{m}}$:
\bea\label{Casimir} 
P_\cG(t; \vec{m})=\prod_{i=1}^{\mathrm{rk}(\cG)} \frac{1}{1-t^{2d_i(\vec{m})}}
\eea
where $d_i(\vec{m})$ are the degrees of the Casimir invariants of $H_{\vec{m}}$. We refer the readers to Appendix A of \cite{Cremonesi:2013lqa} for more details on these classical dressing factors.
\ei
We demonstrate the use of formula \eref{ref_HS} to compute the Coulomb branch Hilbert series of various quiver theories in Section \ref{sec:examples}.

\bibliographystyle{ytphys}
\bibliography{ref}

\end{document}